\documentclass[structabstract]{aa}  
\usepackage{graphicx}
\usepackage{txfonts}
\usepackage{natbib}
\usepackage{color}
\newcommand{\comment}[1]{}
\newcommand{\vel}{\upsilon}

\begin{document}
   \title{Three-dimensional simulations of near-surface convection in main-sequence stars}
   \subtitle{III. The structure of small-scale magnetic flux concentrations}

   \author{B. Beeck\inst{1}
          \and
	  M. Sch\"ussler\inst{1}
	  \and
          R.~H. Cameron\inst{1}
     	  \and
	  A. Reiners\inst{2}
          }

   \institute{Max-Planck-Institut f\"ur Sonnensystemforschung,
              Justus-von-Liebig-Weg 3, 37077 G\"ottingen, Germany\\
         \and
              Institut f\"ur Astrophysik, 
	      Universit\"at G\"ottingen, 
	      Friedrich-Hund-Platz 1,
	      37077 G\"ottingen, Germany\\
             }

   \date{Received 02 February 2015 / Accepted 23 April 2015}
   \titlerunning{3D simulations of near-surface convection in main-sequence
     stars. III. }

\abstract %
{The convective envelopes of cool main-sequence stars harbour magnetic fields
  with a complex global and local structure. These fields affect the
  near-surface convection and the outer stellar atmospheres in many
  ways and are responsible for the observable magnetic activity of
  stars.}
{Our aim is to understand the local structure in unipolar regions with moderate
  average magnetic flux density. These correspond to plage regions covering a substantial fraction of the surface of the Sun (and likely also the surface of other Sun-like stars) during periods of high magnetic activity.}
{We analyse the results of 18 local-box magnetohydrodynamics simulations covering the upper
  layers of the convection zones and the photospheres of cool main-sequence
  stars of spectral types F to early M. The average vertical field in these simulations ranges from 20 to 500\,G.}
{We find a substantial variation of the properties of the surface
  magnetoconvection between main-sequence stars of different spectral types. As
  a consequence of a reduced efficiency of the convective collapse of flux
  tubes, M dwarfs lack bright magnetic structures in unipolar regions of
  moderate field strength. The spatial correlation between velocity and the
  magnetic field as well as the lifetime of magnetic structures and their
  sizes relative to the granules vary significantly along the model sequence of stellar types.}
{}
%

   \maketitle
%
\section{Introduction}
Cool main-sequence stars have outer convection zones (spectral types F to
about M4) or are fully convective (spectral types later than M4). The turbulent
flows in the electrically conducting plasma along with the stellar
(differential) rotation support a hydromagnetic dynamo that constantly
generates magnetic flux \citep[see][for a review]{Charbonneau10}. A fraction
of this magnetic flux becomes buoyant and reaches the near-surface layers
where it is continuously restructured by the convective flows \citep[for
  reviews, see][]{Fan09,Stein12,Manfred13}. At the same time, the magnetic
field strongly affects the stellar near-surface layers. As the solar example
illustrates, the magnetic field causes a variety of activity phenomena
\citep[see][for a review]{sasma}. While many of these (e.\,g. flares and
coronal mass ejections) are located in the higher atmospheric layers and in
the interplanetary space, the solar magnetic field also has an impact on the
photosphere and the convection zone. Sunspots, pores, and faculae are
well-studied photospheric phenomena that are caused by strong (1.5--3\,kG)
localised magnetic fields near the optical surface. Here, the magnetic field
modifies the radiative and convective properties of the near-surface layers
\citep[see e.\,g.][]{sim_sunspots}.\par
%
Like the Sun, many stars exhibit magnetic activity in the form of flares,
X-ray emission, chromospheric emission lines,
etc. \citep{sasma}. Lightcurve variability owing to spatial brightness
variations (e.\,g. starspots) is nowadays detected in essentially every cool
star \citep[e.\,g.][]{Timo}. Measurements of activity indicators (such as
X-ray luminosity or the S-index) reveal that rapidly rotating (young) stars
are much more magnetically active than the Sun \citep[][and references
  therein]{Ansgar2014}. This also affects the detection of planetary
companions around these stars \citep[e.\,g.][]{Jeffersetal2013} and probably
also reduces the chance for habitability of the hosted planets
\citep[e.\,g.][]{Vidottoetal2013}.\par
%
For stars other than the Sun, current observations cannot directly spatially
resolve activity-related phenomena in the photospheres. While most activity
studies rely on activity indicators, for some stars it is possible to measure
the surface magnetic field with spectroscopic and spectropolarimetric methods
\citep[see][for a review]{Ansgar12}. For cool main-sequence stars, these
measurements yield global magnetic field strengths ranging from a few Gauss
\citep[from spectropolarimetry, e.\,g.][]{Petit08} up to $\gtrsim
3\,\mathrm{kG}$ \citep{ReinersBasri2007}. However, it is known from the Sun
that the magnetic field at the surface is highly structured and inhomogeneous
on length scales down to the current resolution limit of solar observations
($\sim 100\,\mathrm{km}$) and probably below. This local structure inevitably
has a significant effect on spectroscopic and spectropolarimetric data \citep{Stenflo13}. In
order to better understand the global magnetic field properties, knowledge of
the local structure of the magnetised photosphere is therefore essential.\par
%
Since the 1990s, comprehensive 3D radiative magnetohydrodynamics (MHD) simulations have been used to
simulate various magnetic phenomena in the solar near-surface layers, ranging
from simple network and plage regions \citep{NS90,MURaM2} to pores
\citep{sim_pores}, sunspots \citep{sim_sunspots}, active regions
\citep{sim_AR1}, emergence of horizontal magnetic flux \citep{NS06,sim_AR2}, and
small-scale dynamo action \citep{Vog07,sim_SSD}. Until a few years ago, MHD
simulations of this kind were only available for the Sun. Recently, the first
comprehensive 3D radiative MHD simulations for stars other than the Sun were
presented \citep{CS16,Wedemeyeretal13,Steineretal13,Steineretal14,CS18}.\par
%
In \citet[][hereafter Paper~I and Paper~II]{paper1,paper2}, we discussed a
sequence of non-magnetic (hydrodynamic) 3D simulations of near-surface
convection with parameters close to main-sequence stars of spectral types
ranging from F3 to M2. In Paper~I, we analysed the overall structure of the
convection and horizontally averaged stratifications. In Paper~II, the
granulation pattern in computed intensity images was studied and the effect of
the 3D nature of the convection on limb darkening and on the profile of
spectral lines was analysed. Here we continue this series of papers by considering
the effects of a moderate magnetic field on the near-surface convection in
cool main-sequence stars. For each set of stellar parameters presented in
Paper~I and Paper~II, we consider three MHD simulation
runs with a unipolar average magnetic field of various strengths. A brief
description of the simulation setup is outlined in Sect.~\ref{sec:setup}. In
Sect.~\ref{sec:magstr}, the distribution of the magnetic field near the
optical surface is discussed. In Sects.~\ref{sec:fluxcon} and~\ref{sec:pr_ba},
we study the structure of small-scale magnetic flux concentrations. In the
following sections, the effect of the magnetic field on the heating of the
upper photosphere (Sect.~\ref{sec:up_ph}) and on the convective flows
(Sect.~\ref{sec:vel}) is analysed. A brief conclusion follows in
Sect.~\ref{sec:disc}.\par
In a subsequent paper of the series \citep[herafter Paper~IV]{paper4},
we will consider the effects of the magnetic field on the limb darkening and
on spectral lines.
\section{Simulation setup}\label{sec:setup}

The simulations discussed in this paper have been obtained with the 3D
radiative MHD code \texttt{MURaM} \citep{MURaM1,MURaM2}. The version of the
code used for the simulations considered here includes the modifications made
by \citet{Rempel09} to enhance the efficiency of the magnetic runs. We limit
the Alfv{\'e}n speed to $c_{\max}=60\,\mathrm{km\,s^{-1}}$ to prevent
exceedingly small time steps. This only affects the nearly force-free regions
near the tops of the simulations boxes and does not considerably
influence the results discussed in this paper and in Paper~IV. The numerical
scheme is described in the appendix of \citet{Rempel09}.\par 
The radiative transfer module of the code utilises the opacity binning method
(see discussion in Paper~I). For the non-magnetic simulations considered here,
opacity distribution functions provided by the \texttt{ATLAS9} code
\citep{atlas9} were rearranged in four groups applying the $\tau$-sorting
method (bin limits at $\log \tau_{\mathrm{ref}}=0,-2,-4$; see discussion in
Paper~I). The magnetic simulations used the same binned opacities as the
corresponding non-magnetic simulations.\par
Six sets of stellar parameters ($\log g$ and $T_{\mathrm{eff}}$) roughly
matching main-sequence stars of spectral types F3V, G2V, K0V, K5V, M0V, and
M2V were considered. The horizontal and vertical sizes of the computational
domains were scaled with the expected granule sizes and pressure scale
heights, respectively, to obtain similar setups for all sets of stellar
parameters (for details, see Paper~I). Snapshots of the non-magnetic
simulations presented in Paper~I and Paper~II were used as initial
configuration for the mass density, the internal energy density, and the
velocity field of the magnetic simulations. The initial magnetic field was
unipolar, vertical with a uniform field strength of $B_0$. For each
non-magnetic simulation, three simulation runs with $B_0=20\,\mathrm{G}$,
$100\,\mathrm{G}$, and $500\,\mathrm{G}$, respectively, and otherwise
identical initial conditions were performed. In a transient restructuring
phase, the flows advect the field into the converging downflow regions. After
some time, the properties of magnetic field and flows become statistically
stationary. The duration of the restructuring phase is of the order of the
local convective turnover time, i.\,e. several minutes at the optical surface
for all stellar parameters of our simulation sequence (and somewhat longer in
the deeper layers). The simulations were run considerably longer to make sure
that a statistically stationary state was reached. Table~\ref{tab:times} gives
the runtime of the simulation $t_1$ (in simulated stellar time) at which the first
snapshots were considered and the total runtime, $t_{\mathrm{end}}$, of the
simulations. The effective temperature, $T_{\mathrm{eff}}$, of the simulations
is regulated only by the entropy density of the inflowing plasma at the bottom
of the computational domain. This quantity was fixed and identical in all four
runs for each set of stellar parameters. Owing to the different (hot and cool)
magnetic small-scale structures, there are slight differences in
$T_{\mathrm{eff}}$ between the different runs of the same star; values are
given in Table~\ref{tab:times}. Values of $T_{\mathrm{eff}}$ of the
non-magnetic runs are almost identical to the ones of the 20\,G runs and are
given in Table~1 of Paper~I. Overall the difference in $T_{\mathrm{eff}}$
between runs of different $B_0$ is small. The reasons for the mostly positive
deviations of the 100\,G and 500\,G runs from the non-magnetic and 20\,G runs
will be discussed in Sect.~\ref{sec:up_ph}.\par
%
\begin{table}
\caption{Stellar parameters and runtime (in simulated minutes) of the simulations.}\label{tab:times}
\centering
\begin{tabular}{lrrrr}\hline\hline
Run (SpT-$B_0$\,$^{\mathrm{a}}$) & $\log g [\mathrm{cgs}]$ & $T_{\mathrm{eff}} [\mathrm{K}]$ & $t_1$\,$^{\mathrm{b}}$ & $t_{\mathrm{end}}$\,$^{\mathrm{c}}$\\\hline
F3V-20G & 4.301 & $6885\pm 6$ & 111 & 127\\
F3V-100G & 4.301 & $6911 \pm 8$ & 106 & 121\\
F3V-500G & 4.301 & $7003 \pm 5$ & 116 & 133\\\hline
G2V-20G & 4.438 & $5779 \pm 9$ & 115 & 133\\
G2V-100G & 4.438 & $5802 \pm 8$ & 108 & 126\\
G2V-500G & 4.438 & $5864 \pm 9$ & 108 & 126\\\hline
K0V-20G & 4.609 & $4858 \pm 2$ & 102 & 117\\
K0V-100G & 4.609 & $4878 \pm 4$ & 78 & 91\\
K0V-500G & 4.609 & $4901 \pm 2$ & 74 & 87\\\hline
K5V-20G & 4.699 & $4376 \pm 2$ & 72 & 83\\
K5V-100G & 4.699 & $4383 \pm 3$ & 61 & 73\\
K5V-500G & 4.699 & $4402 \pm 2$ & 60 & 72\\\hline
M0V-20G & 4.826 & $3907 \pm 1$ & 83 & 96\\
M0V-100G & 4.826 & $3909 \pm 1$ & 58 & 69\\
M0V-500G & 4.826 & $3906 \pm 1$ & 51 & 60\\\hline
M2V-20G & 4.826 & $3691 \pm 1$ & 107 & 124\\
M2V-100G & 4.826 & $3692 \pm 1$ & 59 & 68\\
M2V-500G & 4.826 & $3679 \pm 1$ & 48 & 56\\\hline
\end{tabular}
\begin{list}{}{}
\item[$^{\mathrm{a}}$] SpT: spectral type (of corresponding non-magnetic run); $B_0$: average vertical magnetic field strength
\item[$^{\mathrm{b}}$] $t_1$: first time step considered in the analysis
\item[$^{\mathrm{c}}$] $t_{\mathrm{end}}$: current end time of simulation
\end{list}
\end{table}
The combination of three different values for $B_0$ with the six different
sets of stellar parameters results in 18 magnetic simulations.\footnote{Movies
  of the vertical velocity and the magnetic field strength at the optical
  surface as well as the bolometric intensity of the 18 magnetic simulations
  are provided as online material. They show the temporal evolution of
  the granulation during the time interval considered in this paper and
  Paper~IV.} The six non-magnetic simulations presented in Paper~I and
Paper~II serve as a reference, completing the grid of 24 simulation runs.

\section{Magnetic field distribution at the optical surface}\label{sec:magstr}
\begin{figure*}
\centering
  \includegraphics[width=4.98cm]{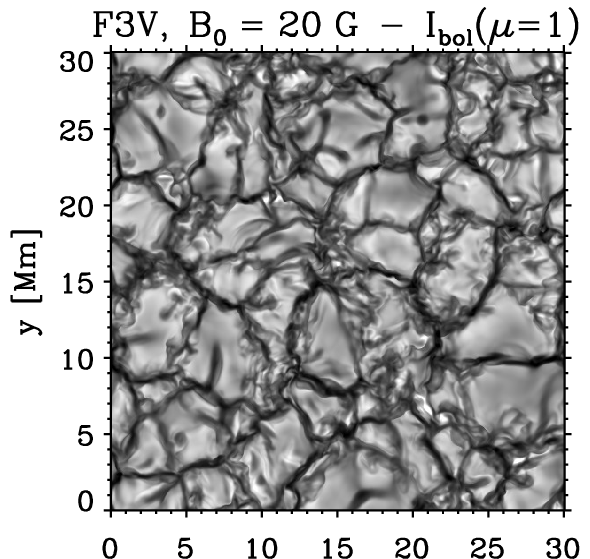}~\includegraphics[width=4.62cm]{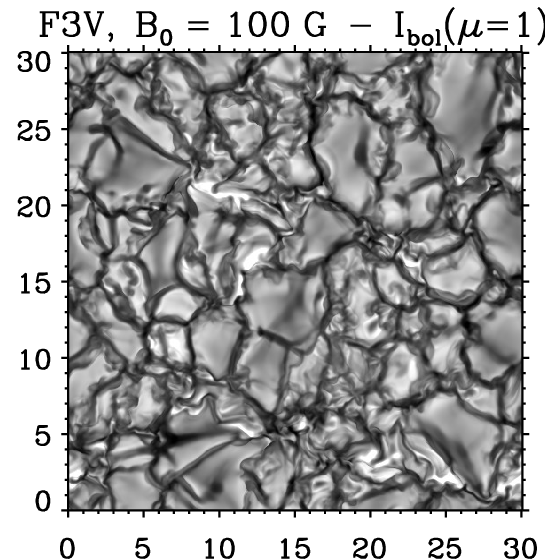}~\includegraphics[width=4.62cm]{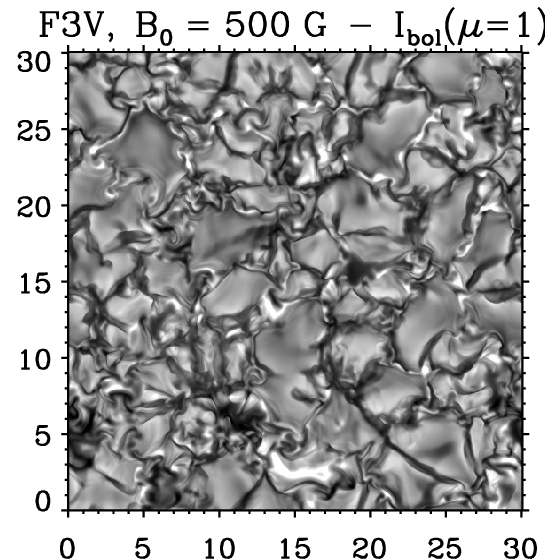}~\includegraphics[width=1.41cm]{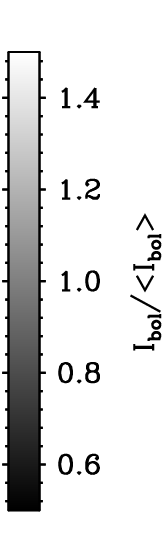}\\
  \includegraphics[width=4.98cm]{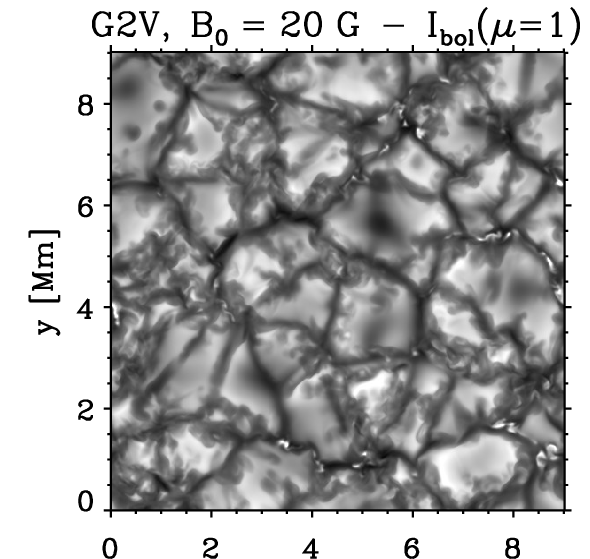}~\includegraphics[width=4.62cm]{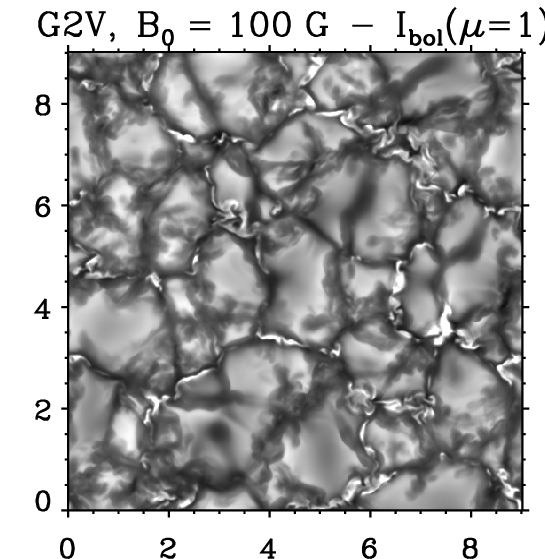}~\includegraphics[width=4.62cm]{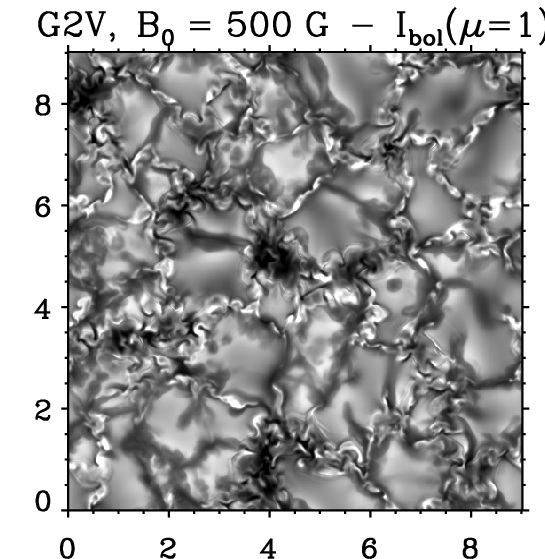}~\includegraphics[width=1.41cm]{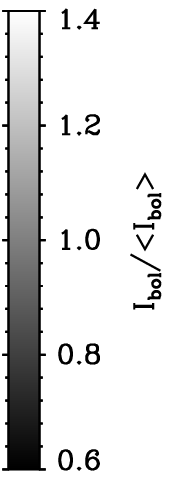}\\
  \includegraphics[width=4.98cm]{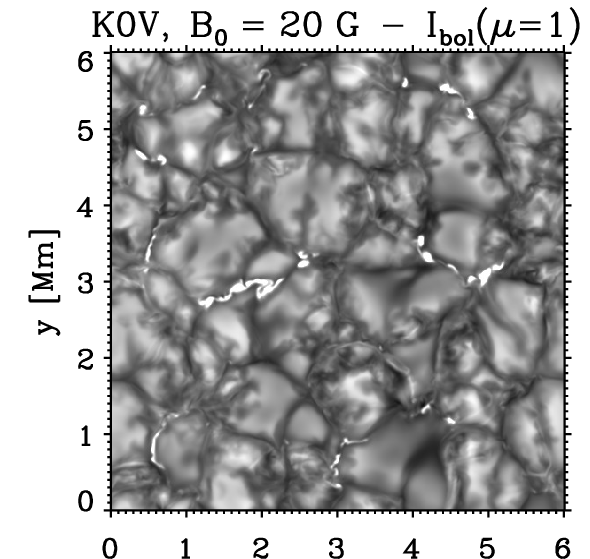}~\includegraphics[width=4.62cm]{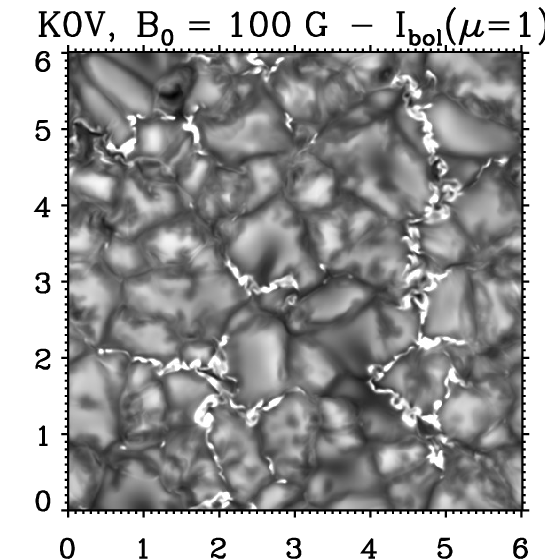}~\includegraphics[width=4.62cm]{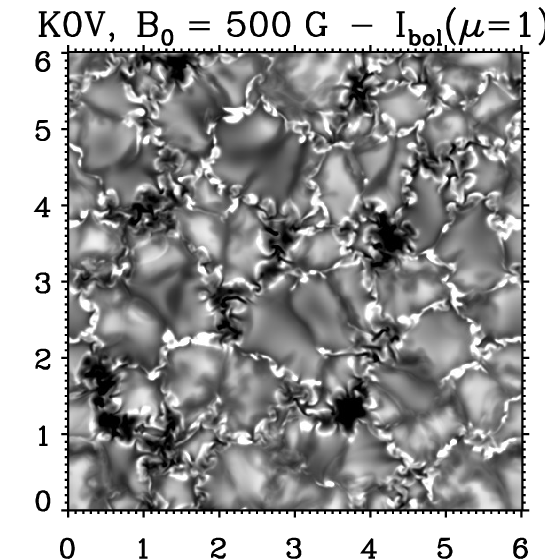}~\includegraphics[width=1.41cm]{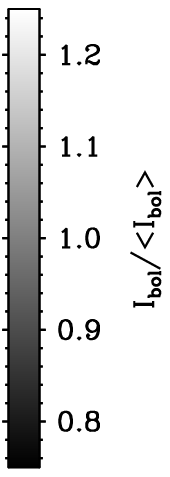}\\
  \includegraphics[width=4.98cm]{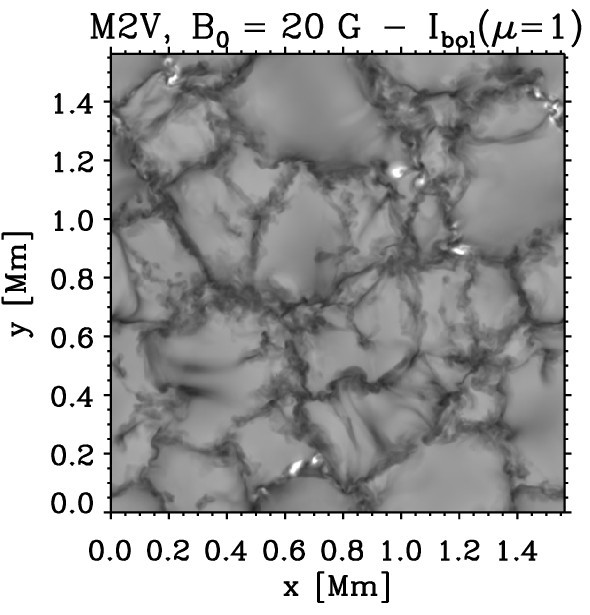}~\includegraphics[width=4.62cm]{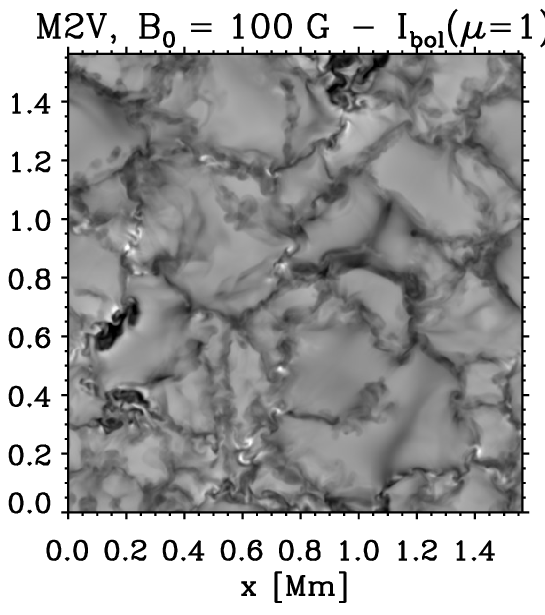}~\includegraphics[width=4.62cm]{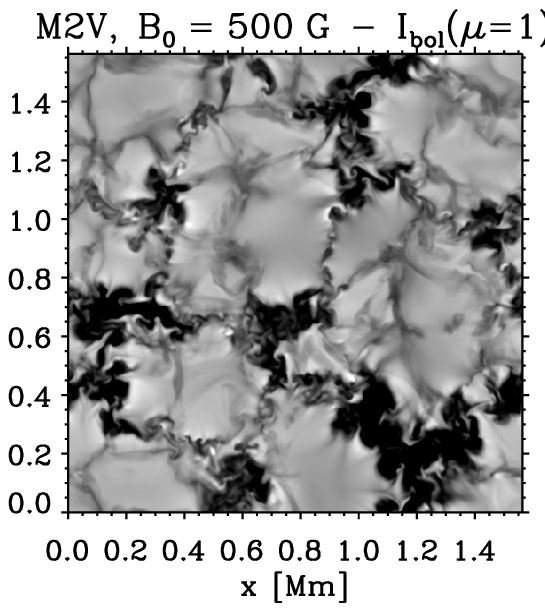}~\includegraphics[width=1.41cm]{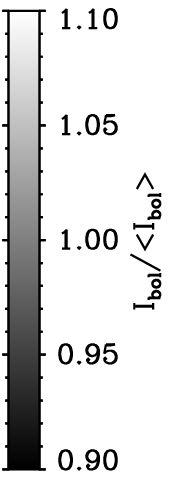}\\
\caption{Maps of the vertical bolometric intensity for twelve of the 18
  magnetic simulation runs. For improved image contrast, the grey scale saturates at the values indicated by the colour scales on the right of each row.}\label{fig:mag_int_1}
\end{figure*}
\begin{figure*}
  \centering
  \includegraphics[width=4.75cm]{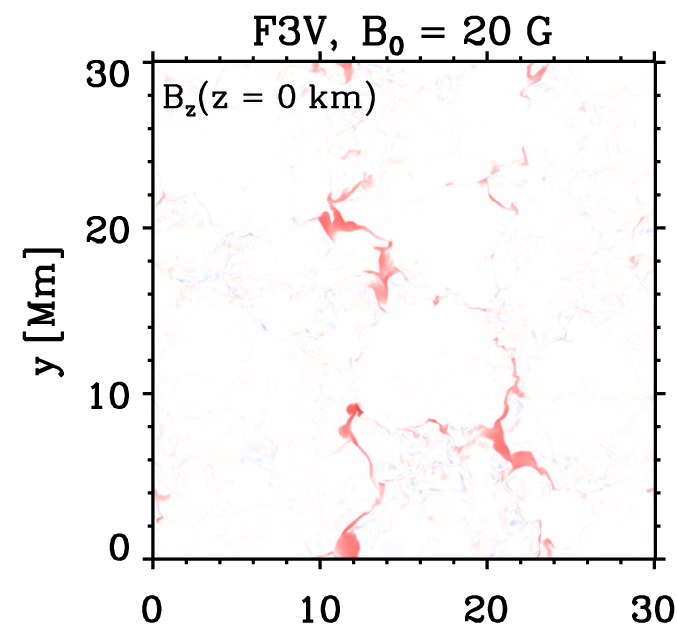}\includegraphics[width=4.42cm]{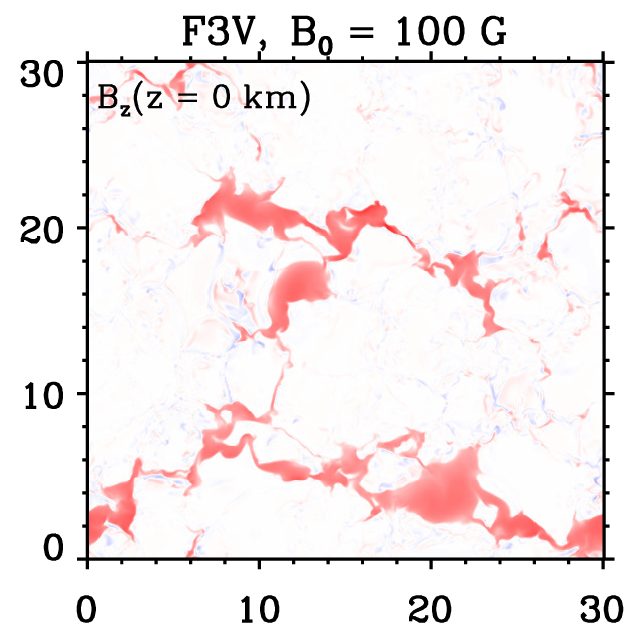}\includegraphics[width=4.42cm]{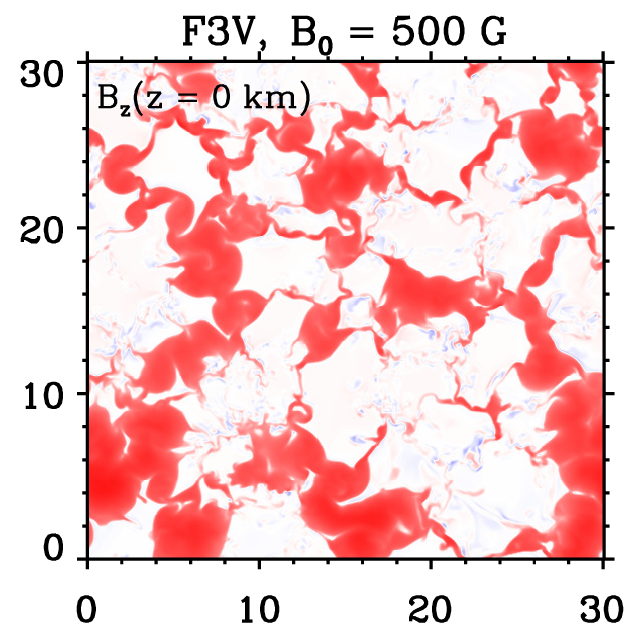}\\
  \includegraphics[width=4.75cm]{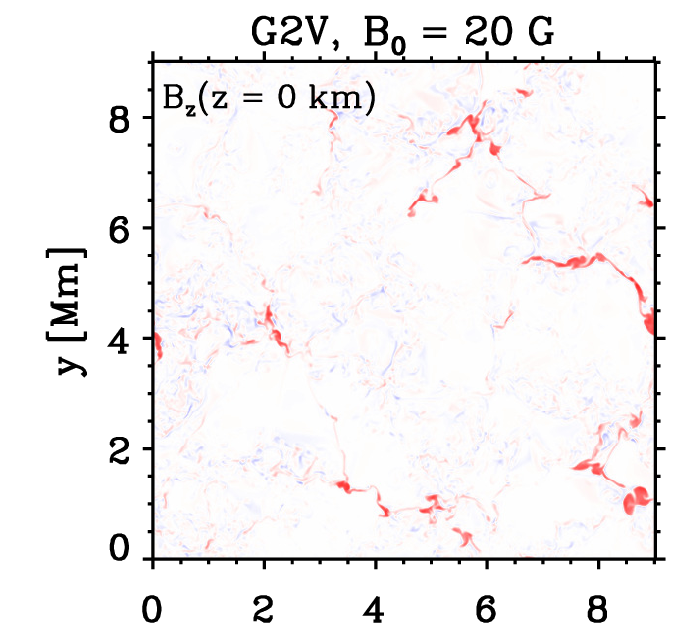}\includegraphics[width=4.42cm]{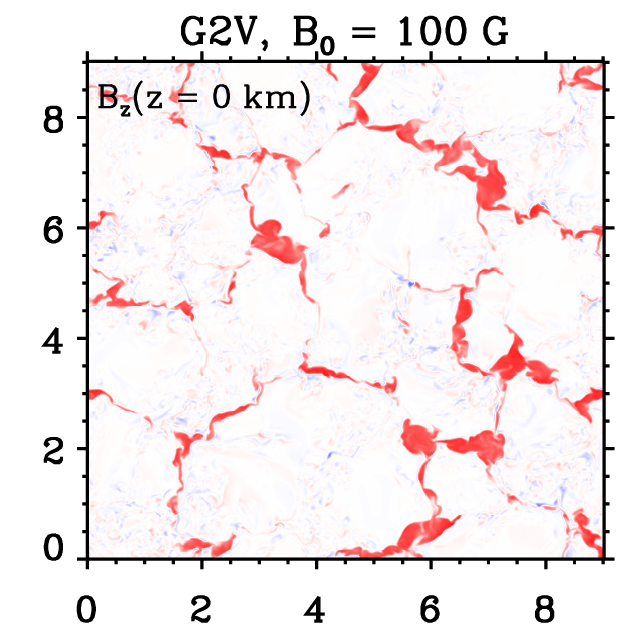}\includegraphics[width=4.42cm]{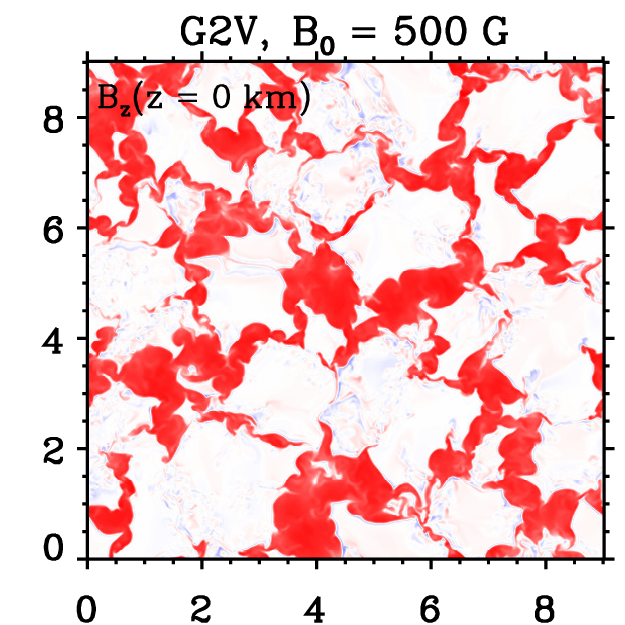}\\
  \includegraphics[width=4.75cm]{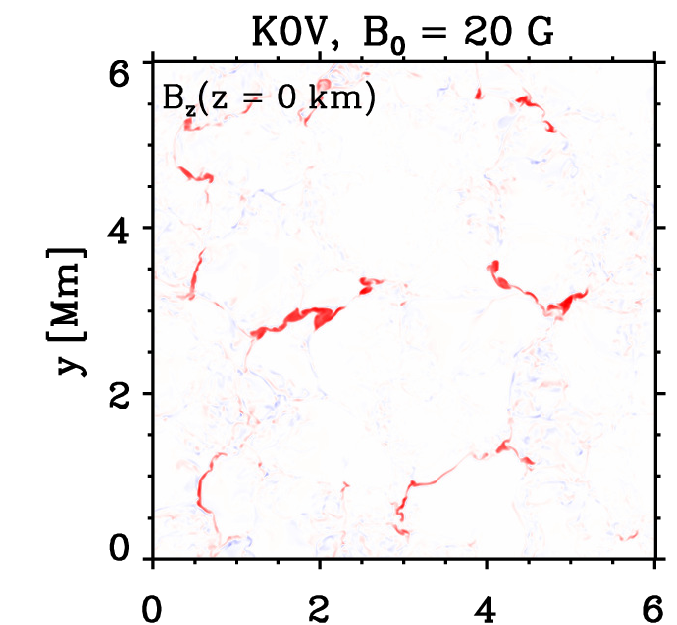}\includegraphics[width=4.42cm]{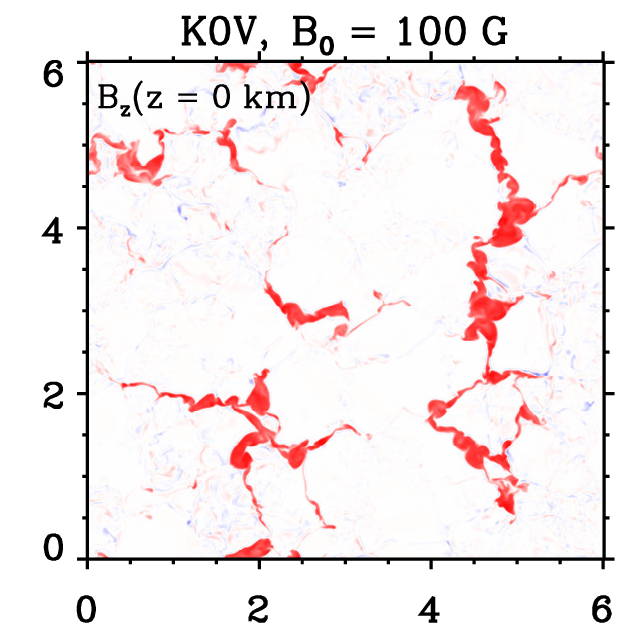}\includegraphics[width=4.42cm]{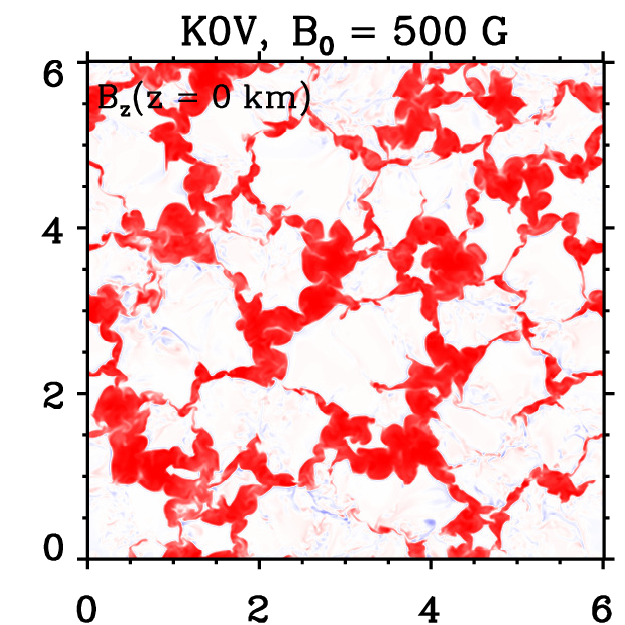}\\
  \includegraphics[width=4.75cm]{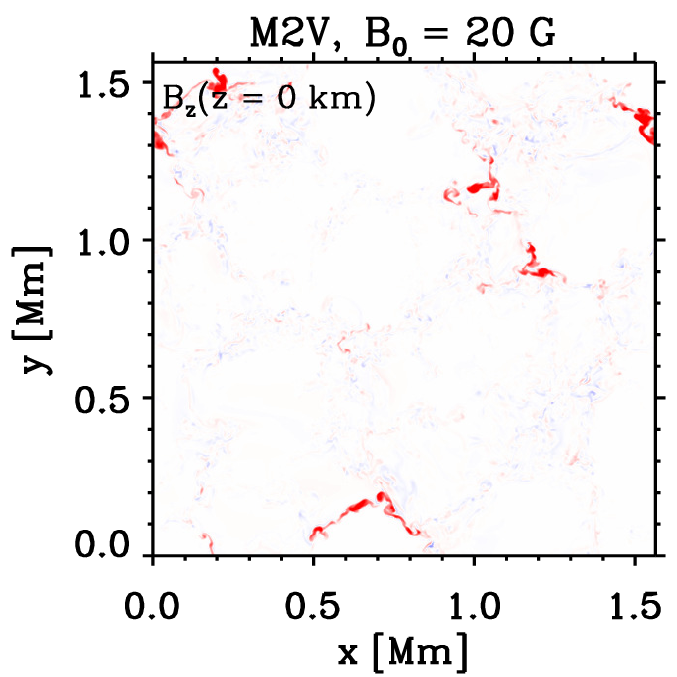}\includegraphics[width=4.42cm]{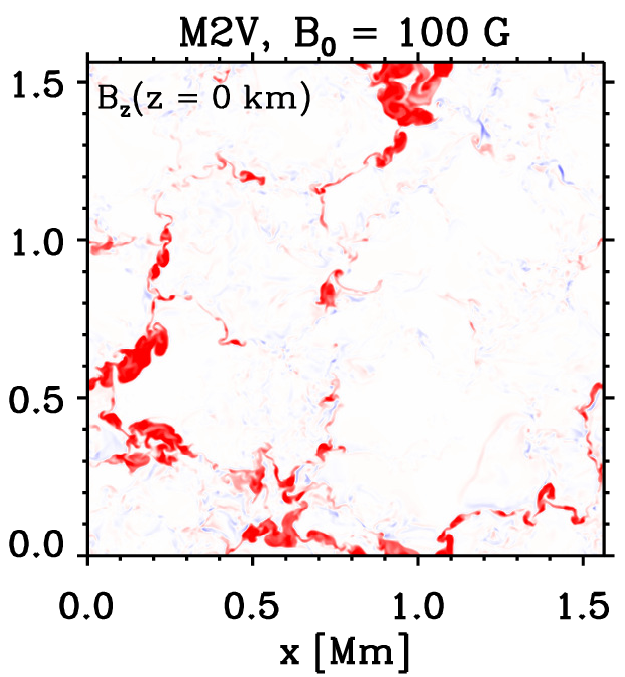}\includegraphics[width=4.42cm]{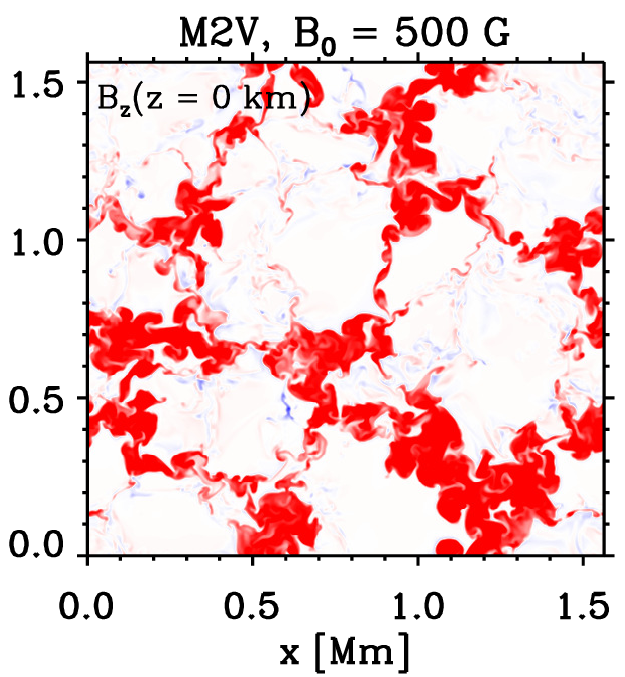}\\
  \includegraphics[width=13.3cm]{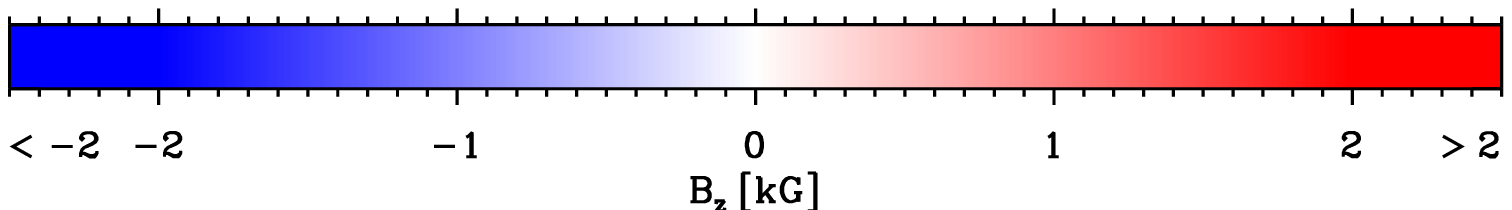}
\caption{Maps of the vertical component, $B_z$, of the magnetic field at $z=0$, the average geometrical depth of the optical surface (same simulation runs and at the same time step as in Fig.~\ref{fig:mag_int_1}).}\label{fig:magmap1}
\end{figure*}

Figure~\ref{fig:mag_int_1} shows the vertically emerging bolometric intensity
of all three magnetic runs for the four spectral types F3V, G2V, K0V, and
M2V. An analogous figure for the non-magnetic simulations is given in Fig.~2
of Paper~I. In all simulations, regions of either strongly enhanced or
significantly reduced intensity appear, mostly situated in the intergranular
lanes. These regions coincide with the locations of strong magnetic flux
concentrations (see~Fig.~\ref{fig:magmap1}). The relative number and size of
dark and bright regions as well as their intensity contrast strongly depend on
the spectral type and the amount of magnetic flux in the simulation box
(i.\,e. on $B_0$). This is discussed in some more detail in
Sect.~\ref{sec:fluxcon}. The most striking difference appears between the
M-star simulations on the one hand and all other simulations on the other: in
the F-, G- and K-stars simulations, only for $B_0=500\,\mathrm{G}$ a few dark
features appear among the mostly bright magnetic structures; in contrast, in
the M2V star (and in the M0V star; not shown here), dark magnetic
structures frequently form already in the 100\,G runs and the intensity maps of
the 500\,G runs are dominated by large dark structures while there are
essentially no bright magnetic structures left \citep[cf.][]{CS16}.\par
%
\begin{table}
\caption{Properties of the magnetic field at the $\tau_{\mathrm{R}}=1$ surface (all values in G; errors give temporal 1-$\sigma$ scatter)}\label{tab:mag_val}
\centering
\begin{tabular}{crr@{$\,\pm\,$}lr@{$\,\pm\,$}lr@{$\,\pm\,$}l}
\hline\hline
SpT\,$^{\mathrm{a}}$ & \multicolumn{1}{c}{$\langle B_z\rangle$} & \multicolumn{2}{c}{$\max(B(\tau_{\mathrm{R}}=1))$} & \multicolumn{2}{c}{$\sqrt{\langle B^2\rangle}$} & \multicolumn{2}{c}{$B_{\mathrm{strong}}\,^{\mathrm{b}}$} \\\hline
F3V & 20  & $2040$ & $160$ & $150$ & $4$   & \multicolumn{1}{c@{---}}{} & \\
F3V & 100 & $4250$ & $550$ & $445$ & $4$   & $1418$ & $67$\\
F3V & 500 & $5180$ & $510$ & $1160$ & $10$ & $1886$ & $59$\\\hline
G2V & 20  & $2310$ & $110$ & $178$ & $6$   & \multicolumn{1}{c@{---}}{} & \\
G2V & 100 & $3250$ & $410$ & $412$ & $10$  & $1702$ & $74$\\
G2V & 500 & $3550$ & $150$ & $1022$ & $11$ & $1990$ & $18$\\\hline
K0V & 20  & $2890$ & $160$ & $175$ & $7$   & \multicolumn{1}{c@{---}}{} & \\
K0V & 100 & $2930$ & $130$ & $425$ & $3$   & $1824$ & $24$\\
K0V & 500 & $3390$ & $140$ & $1031$ & $6$  & $2056$ & $11$\\\hline
K5V & 20  & $2640$ & $170$ & $176$ & $4$   & \multicolumn{1}{c@{---}}{} & \\
K5V & 100 & $2980$ & $80$  & $418$ & $6$   & $1823$ & $48$\\
K5V & 500 & $3550$ & $140$ & $1043$ & $3$  & $2099$ & $13$\\\hline
M0V & 20  & $3400$ & $240$ & $185$ & $5$   & \multicolumn{1}{c@{---}}{} & \\
M0V & 100 & $3900$ & $200$ & $436$ & $4$   & $1917$ & $41$\\
M0V & 500 & $4200$ & $200$ & $1075$ & $3$  & $2326$ & $15$\\\hline
M2V & 20  & $3500$ & $580$ & $182$ & $3$   & \multicolumn{1}{c@{---}}{} & \\
M2V & 100 & $4090$ & $370$ & $440$ & $8$   & $1994$ & $85$\\
M2V & 500 & $4450$ & $320$ & $1067$ & $7$  & $2352$ & $57$\\\hline
\end{tabular}
\begin{list}{}{}
\item[$^{\mathrm{a}}$] spectral type of the simulation
\item[$^{\mathrm{b}}$] peaks of the histograms of Fig.~\ref{fig:maghist}
\end{list}
\end{table}
%
Figure~\ref{fig:magmap1} shows maps of the vertical component $B_z$ of the
magnetic field at $z=0$, the mean height level of the optical surface for the
same selection of simulation runs and at the same time steps as
Fig.~\ref{fig:mag_int_1}. The strong concentration of the magnetic flux in
some of the convective downflows leads to local field strengths of up to
several kG in all stars, even in the case of $B_0=20\,\mathrm{G}$. The maximum
of the field strength at the optical surface depends only weakly on $B_0$,
especially in the cooler stars of the model sequence: for instance, in the M2V
star, it is 3.5\,kG in the 20\,G run and 4.5\,kG in the 500\,G run (more
values given in Table~\ref{tab:mag_val}).\par
\begin{figure*}
\includegraphics[width=6cm]{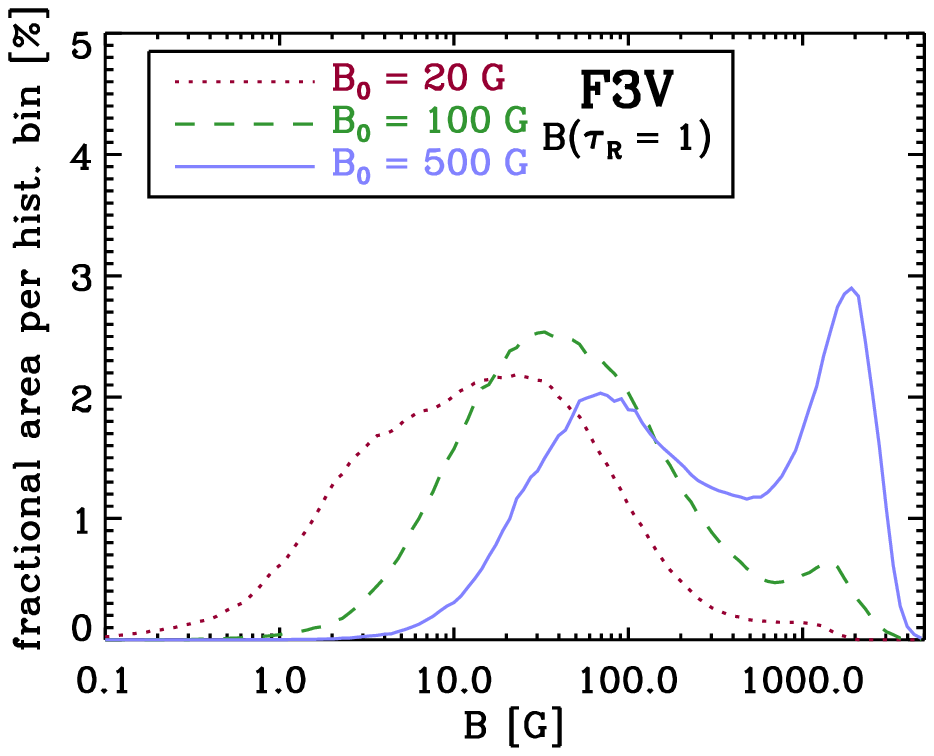}~\includegraphics[width=6cm]{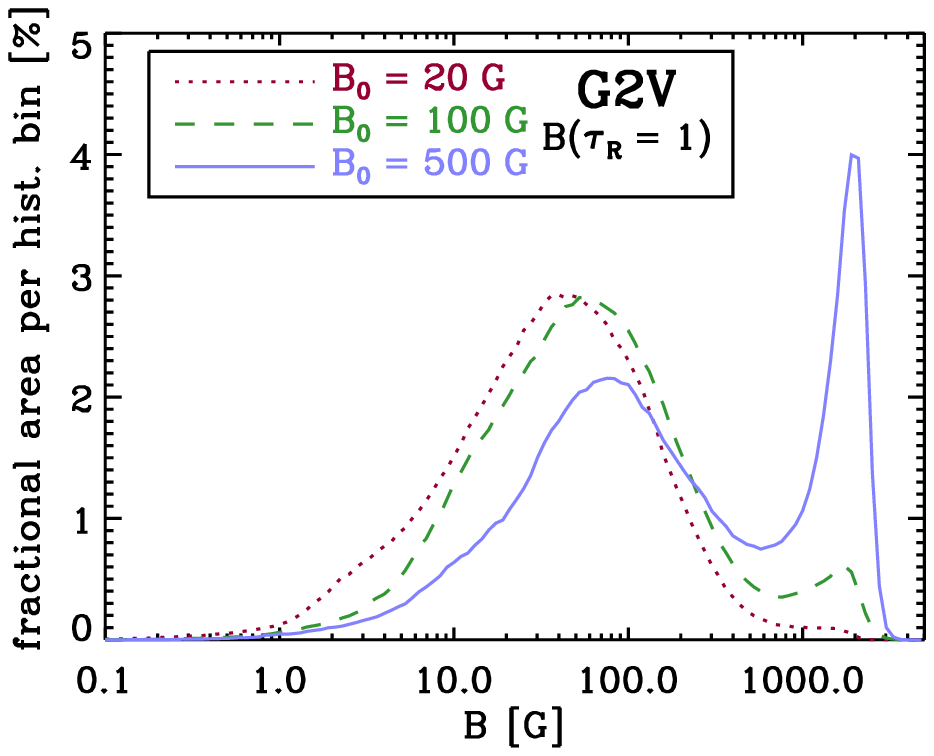}~\includegraphics[width=6cm]{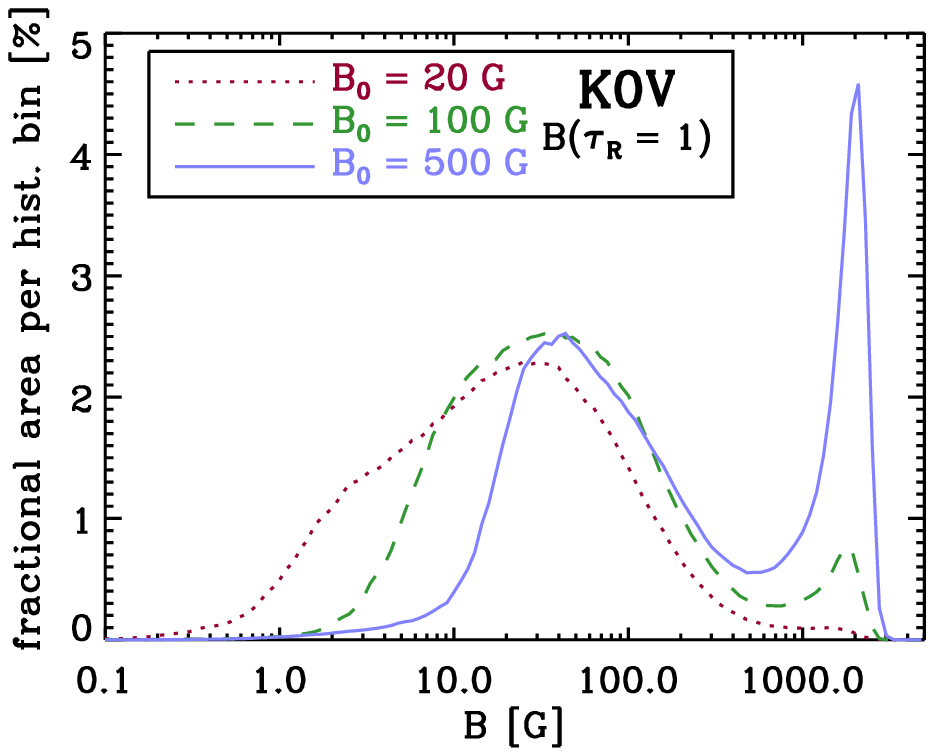}\\
\includegraphics[width=6cm]{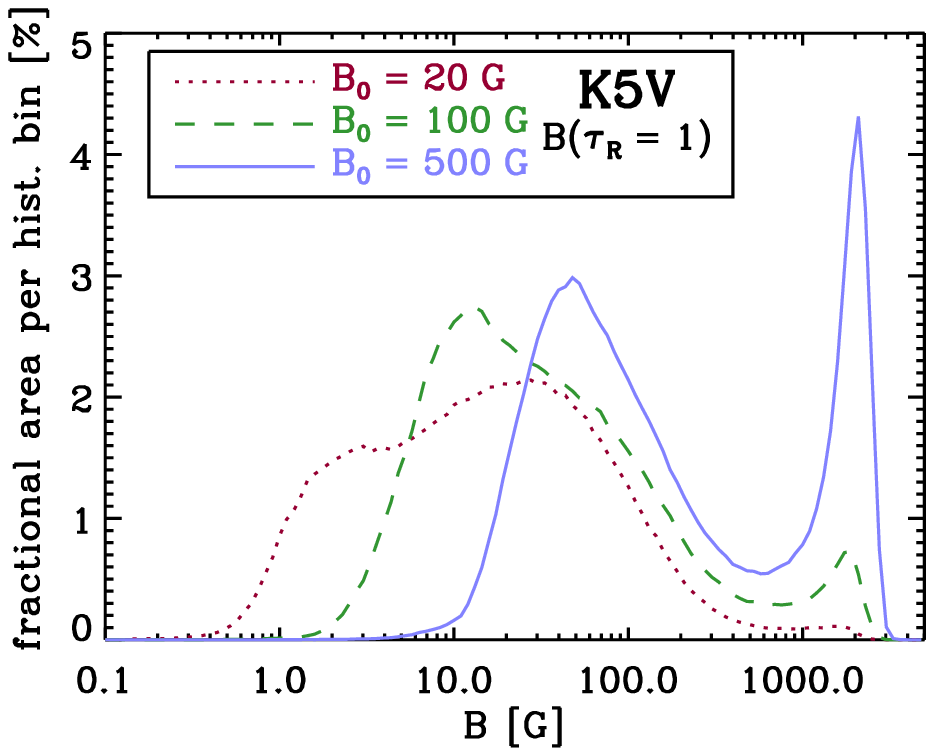}~\includegraphics[width=6cm]{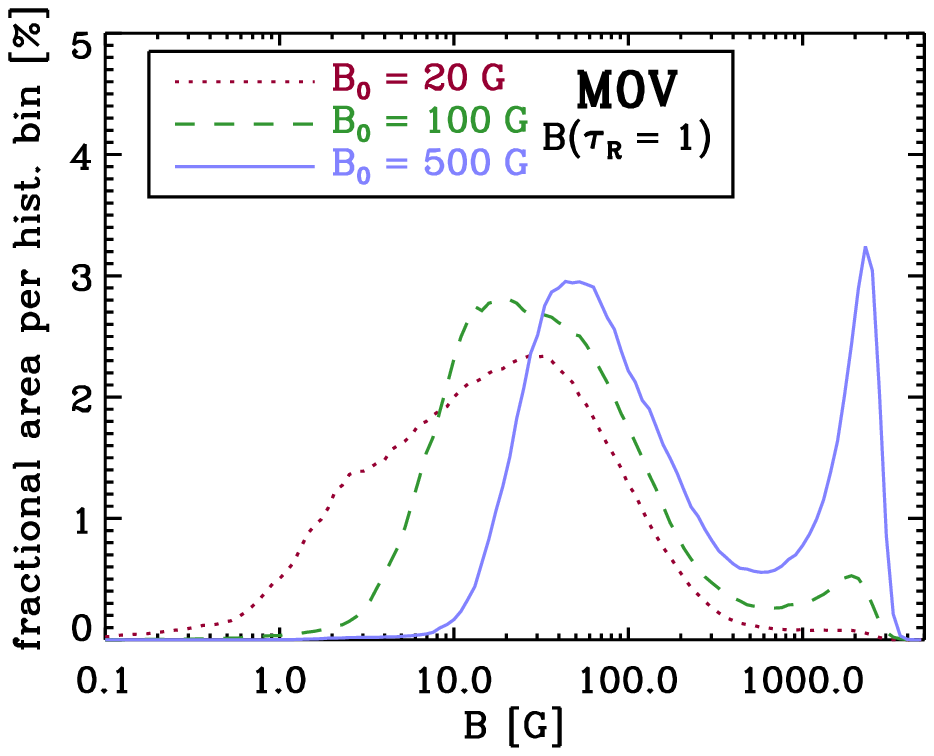}~\includegraphics[width=6cm]{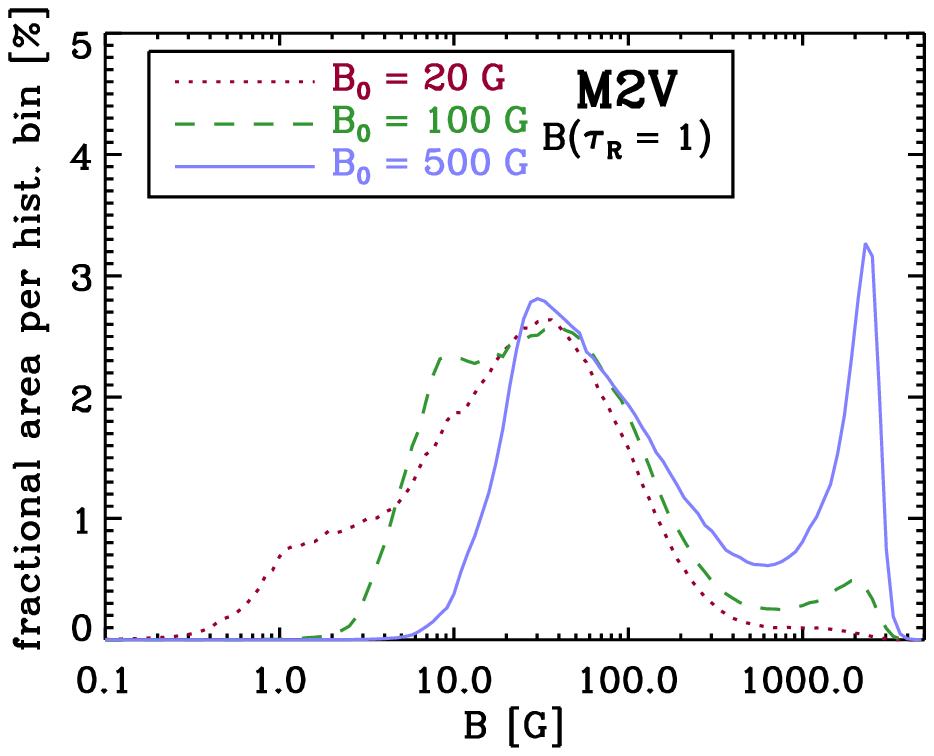}\\
\caption{Histograms of the modulus of the magnetic field strength, $B$, at the
  optical surface for all 18 magnetic simulation runs (the histogram bin size is 0.04\,dex).}\label{fig:maghist}
\end{figure*}
%
As the typical field strength in the magnetic flux concentrations does not
strongly vary with $B_0$, the area fraction of these structures increases with
increasing $B_0$. This is also illustrated by Fig.~\ref{fig:maghist}, which
gives the histograms of the magnetic field modulus, $B(\tau_{\mathrm{R}}=1)$, at
the optical surface. In the 500\,G and 100\,G runs, these histograms have two
peaks (using logarithmically equidistant bins), one at relatively low values
between 10 and 100\,G (in most simulations, this peak falls between 30 and
50\,G) and one at a high field strength of about 2\,kG. The latter peak
represents the typical field strength at the optical surface in the magnetic
flux concentrations and is denoted as $B_{\mathrm{strong}}$ in what
follows. Values for $B_{\mathrm{strong}}$ obtained from a Gaussian fit to the
histogram peak range between 1.4\,kG and 2.4\,kG (see
Table~\ref{tab:mag_val}). $B_{\mathrm{strong}}$ slightly increases with $B_0$
and towards the cooler end of our sequence. In the 20\,G runs there is no strong-field peak, but the distributions of $B(\tau_{\mathrm{R}}=1)$ still show an
extended tail towards high values of the field strength.\par
%
The weak dependence of $B_{\mathrm{strong}}$ on $B_0$ is consistent with the
rms values of $B$, which are largely independent of stellar parameters. This
supports a simple two-component model of the surface field: if one assumes a
fraction $f$ of the stellar surface to have a field strength
$B_{\mathrm{strong}}$ and the remaining fraction $1-f$ to have a field
strength $B_{\mathrm{weak}}\approx 0$ and further assumes the field to be
mostly vertical and unipolar, one obtains $f=B_0/B_{\mathrm{strong}}$, from
which for the rms $\langle B^2\rangle^{1/2}=(f B_{\mathrm{strong}}^2)^{1/2}=
(B_0 B_{\mathrm{strong}})^{1/2}$. For $B_{\mathrm{strong}}=2\,\mathrm{kG}$ one
obtains an rms magnetic field of 200, 447, and 1000\,G for $B_0= 20$, 100, and
500\,G, respectively, which is very close to the values actually found in the
simulations (see Table~\ref{tab:mag_val}). Owing to the strong height
dependence of the magnetic field (cf. Fig.~\ref{fig:WD}), as well as to the
local impact of the magnetic field on the atmosphere structure, a
two-component model appears nonetheless inadequate to compute profiles of
magnetically sensitive spectral lines (cf. Paper~IV).\par

\section{Structure of magnetic flux concentrations}\label{sec:fluxcon}
%
%
\begin{figure*}
\centering
\includegraphics[width=7cm]{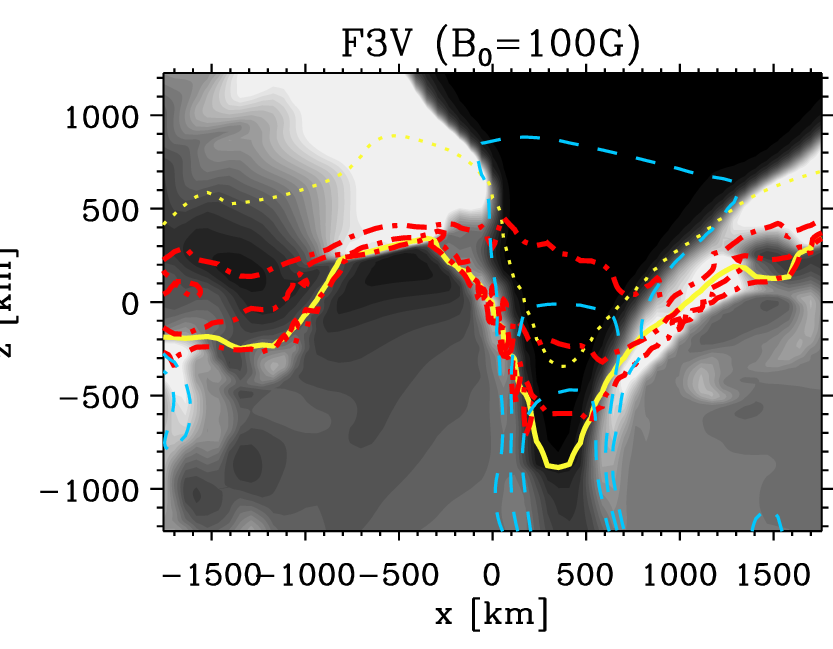}~~~\includegraphics[width=7cm]{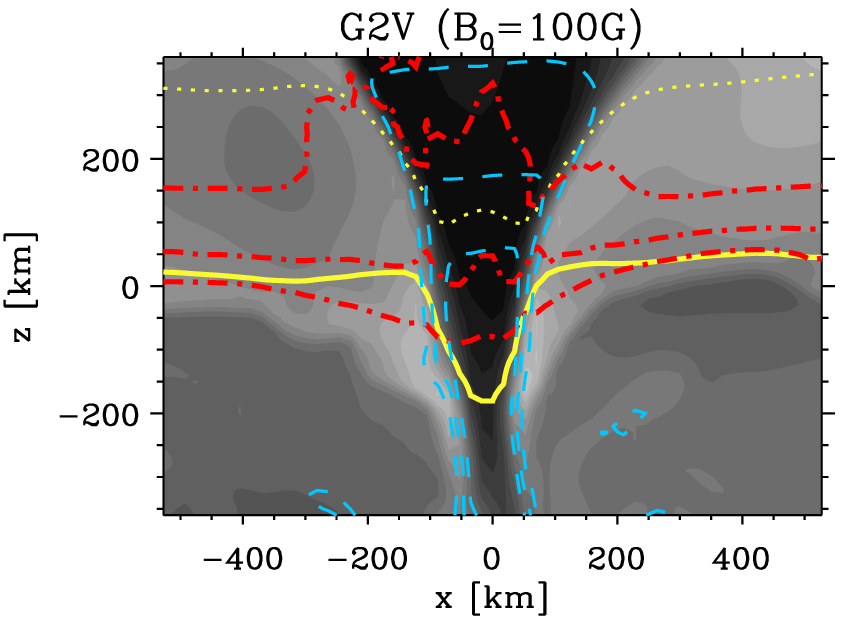}\\
\includegraphics[width=7cm]{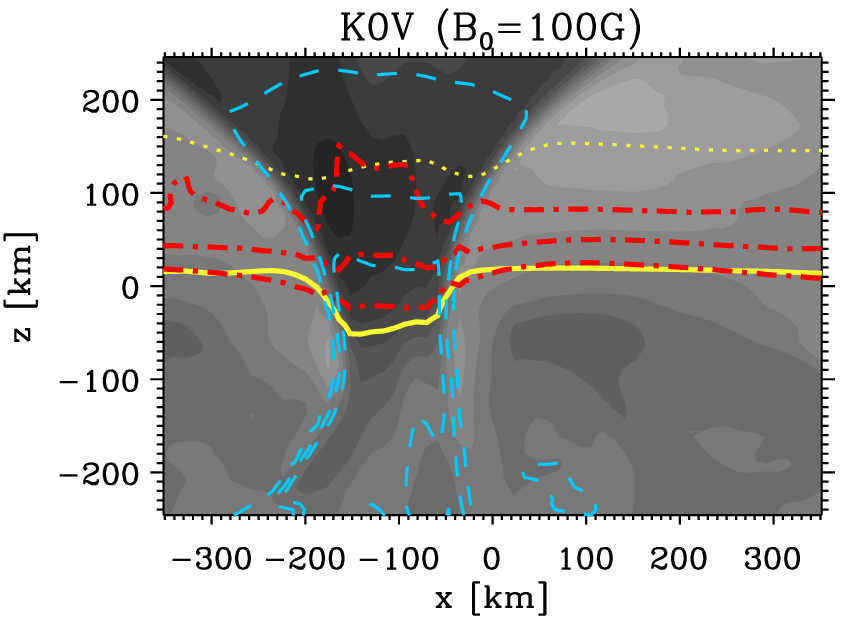}~~~\includegraphics[width=7cm]{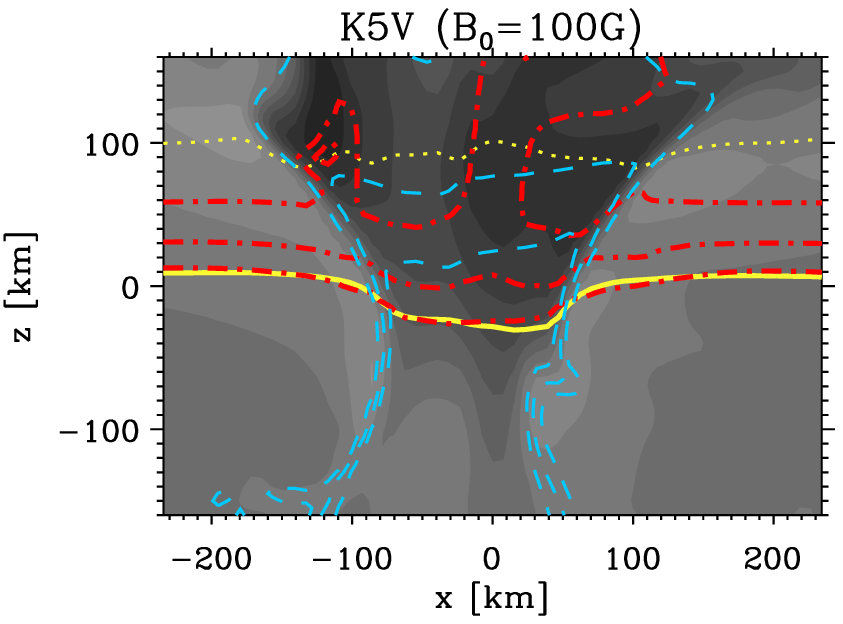}\\
\includegraphics[width=7cm]{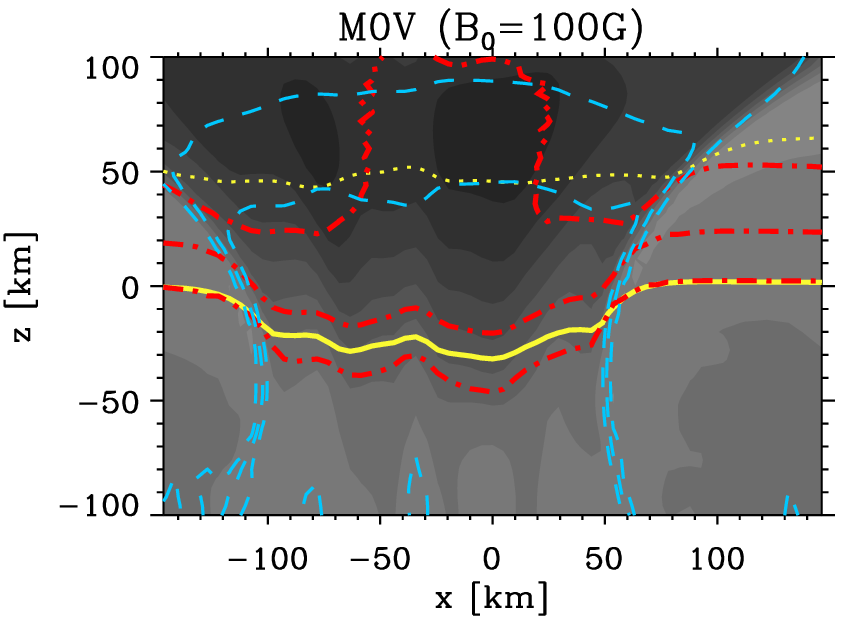}~~~\includegraphics[width=7cm]{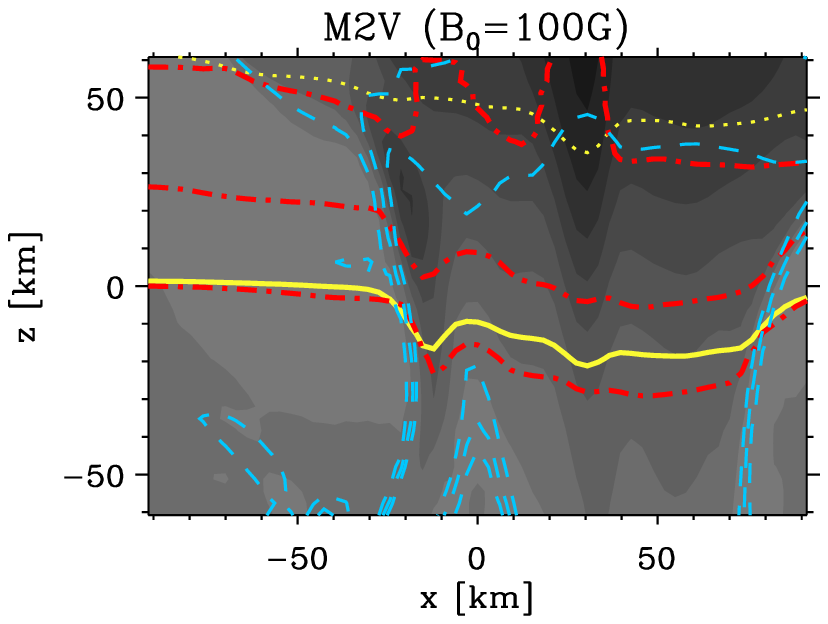}\\
\includegraphics[width=14.3cm]{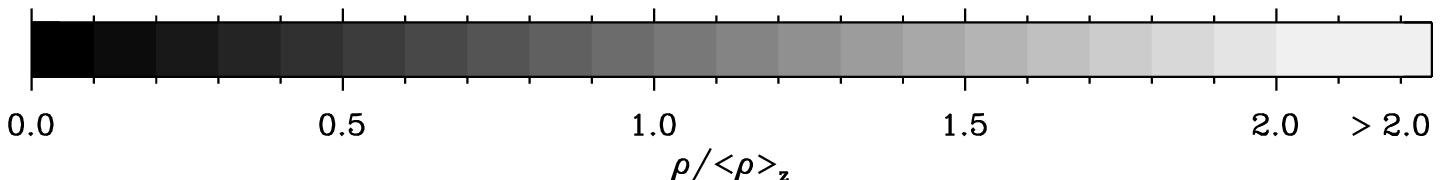}
\caption{Vertical cuts through typical magnetic flux concentrations in the six
  simulations with $B_0=100\,\mathrm{G}$. The grey scale indicates the density
  $\varrho/\langle\varrho\rangle_z$ relative to the horizontal mean density;
  the solid (dotted) yellow curve represents the $\tau_{\mathrm{R}}=1$
  ($\tau_{\mathrm{R}}=0.01$) surface; the dashed blue contours indicate iso-$B$ contours of 500, 1000, and $1500\,\mathrm{G}$; the
  red dash-dotted curves indicate the $T/T_{\mathrm{eff}}=0.9$, $1.0$, and
  $1.1$ surfaces. $x$ and $z$ coordinates are to scale for each panel but
  differ by more than one order of magnitude for the different stellar
  models.}\label{fig:WD}
\end{figure*}
%
Figure~\ref{fig:WD} shows vertical cuts through one of the magnetic flux
concentrations for each of the 100\,G runs. The light-blue dashed curves give
iso-contours of the magnetic field strength of 500, 1000, and 1500\,G, which
indicate the locations of strong magnetic flux concentrations. In the flux
concentrations, the magnetic pressure $p_{\mathrm{mag}}=B^2/(8\pi)$ becomes
substantial. As the atmospheres are roughly in hydrostatic equilibrium, the
gas pressure and the (mass) density are locally reduced in regions of strong
magnetic field. The underlying grey-scale image gives the density distribution
and confirms that the flux concentrations are all under-dense with respect to
the horizontal (iso-$z$) average stratification. However, the degree of
evacuation strongly varies throughout the model sequence. The hottest two
stars (F3V and G2V) show the strongest evacuation in the upper (optically
thin) part of the magnetic structures, with a density of the order of only
10\% of its horizontal mean value (in the first pressure scale height above
the optical surface). In the K-dwarf simulations, the density in the magnetic
structures reaches still about 30-50\% of the horizontal average density and
in the M-dwarf simulations, where evacuation is more confined to the upper
layers, there is almost no evacuation near the optical surface. This effect is
probably a result of the much lower superadiabaticity in the cooler stars (see
Paper~I), which makes the convective collapse less efficient \citep[cf.][]{Rajaguru02}. The reduced efficiency of the convective collapse results in reduced ratio between the magnetic pressure and the gas pressure (see Sect.~\ref{sec:pr_ba}).\par
The solid yellow contours in Fig.~\ref{fig:WD} outline the
$\tau_{\mathrm{R}}=1$ surface (= optical surface). In all simulations,
magnetic flux concentrations cause a local depression in the optical surface,
which is referred to as Wilson depression in analogy to the same phenomenon in
sunspots. The depression is caused by the partial evacuation in the
near-surface layers, which results in a lower absorption coefficient,
$\kappa\varrho$, (and thus longer mean free path length of the photons).\par
The red dash-dotted curves outline iso-$T$ contours of $0.9$, $1.0$ and $1.1$
times the effective temperature. In all cases,
the $1.1\,T_{\mathrm{eff}}$ iso-contour has a lower geometrical height within
the magnetic flux concentrations than outside: at the same {\it geometrical}
depth, all flux concentrations are cool around $z=0$. However, the optical
surface depressions are in most cases larger than the depressions of the iso-$T$
contours: at the same {\it optical} depth, most flux concentrations shown here
are hotter than the surroundings (with the notable exception of the M-star
simulations). The contours of $1.0$ and $0.9\,T_{\mathrm{eff}}$ are mostly above the optical
surface and show a more complicated structure.\par
The optically thin plasma in the photospheric layers of the flux
concentrations is heated by various mechanisms including adiabatic heating in
downflows (subadiabatic temperature gradient), radiative heating, as well as viscous
and Ohmic dissipation (see Sect.~\ref{sec:up_ph}). In the layers directly
above the optical surface, the absorption coefficient is still high enough for
radiative heating to be important. This leads to heating near inclined and
concavely curved parts of the optical surface (side walls of the
depressions). As a consequence, magnetic flux concentrations with sizes comparable to the Wilson depression have higher temperatures than the non-magnetic
surroundings near the optical surface. In flux
concentrations with a large diameter (compared to their depth), this heating is
only effective near the side walls. This explains why in the F-, G-, and
K-type simulations the larger magnetic flux concentrations become dark in
their centres while smaller ones are bright. In the M-star simulations, the
depressions in the optical surface are all very shallow owing to the low
degree of evacuation in the upper part of the flux
concentration. Consequently, even a moderate amount of magnetic flux generates
a structure with a large radius compared to its Wilson depression. Therefore,
many magnetic flux concentrations on M stars are dark.\par
The dotted yellow contours in Fig.~\ref{fig:WD} give the $\tau_{\mathrm{R}}=0.01$ surface. Many
photospheric spectral lines (such as the ones discussed in Paper~II and
Paper~IV) are most sensitive to the conditions around this optical
depth. The magnetic field strength in the
flux concentrations and their filling factor may differ by a significant amount
between the optical surface and this higher iso-$\tau_{\mathrm{R}}$
surface. Note that the depressions in the $\tau_{\mathrm{R}}=0.01$ surface are
generally shallower than the depressions of the optical surface. This can
partly be attributed to the heating mechanisms, which lead to a positive
temperature deviation (with respect to the horizontal average) at this height (and above) and thus to a
higher opacity, partially compensating the diminishing effect of the reduced
density on the absorption coefficient.\par
%
\begin{figure*}
\centering
\includegraphics[width=7.1cm]{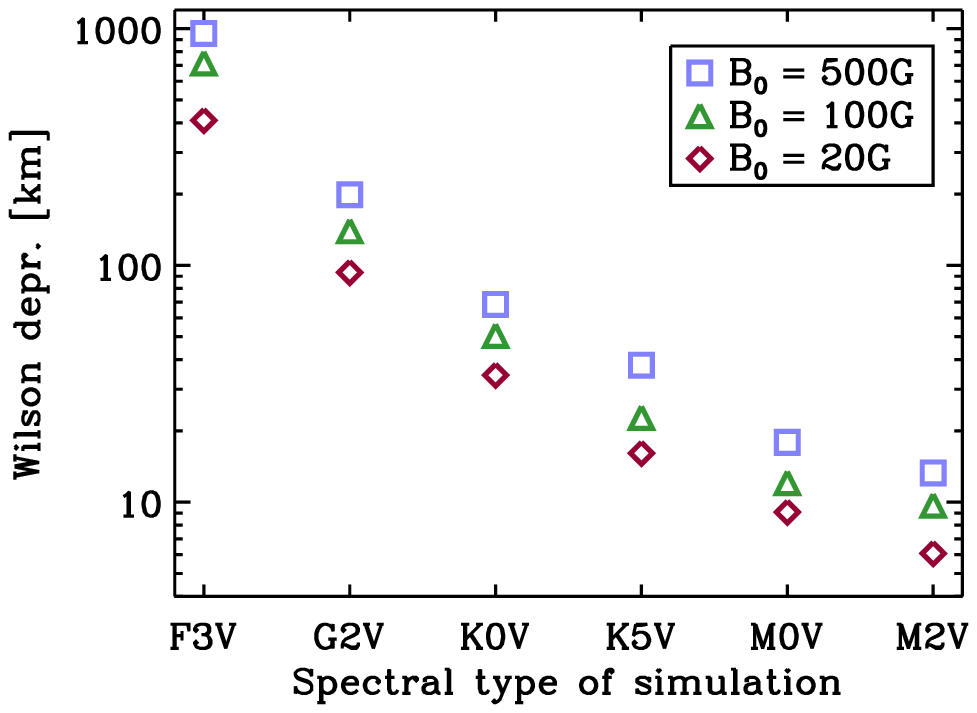}~\includegraphics[width=7.1cm]{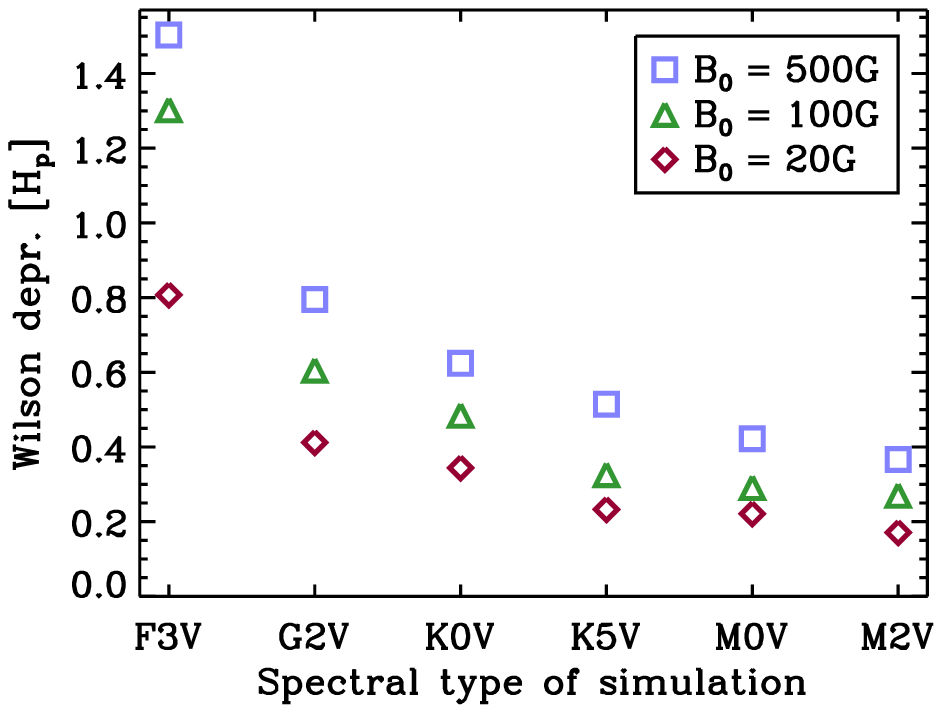}
\caption{Wilson depressions. {\it Left panel:} Difference in average geometrical depth of the $\tau_{\mathrm{R}}=1$ surface between magnetic ($B(\tau_{\mathrm{R}}=1)>750\,\mathrm{G}$) and non-magnetic area ($B(\tau_{\mathrm{R}}=1)<750\,\mathrm{G}$) as a measure for the depth of the Wilson depressions. {\it Right panel:} Downward shift of the optical surface within the magnetic flux concentrations in terms of pressure scale height (measured outside the flux concentrations).}\label{fig:WDall}
\end{figure*}
%
Figure~\ref{fig:WDall} shows the dependence of the depth Wilson depression on
the spectral type of the simulation and on $B_0$.  In the left panel of the
figure, the depth is given in terms of km. By far the deepest Wilson
depressions are observed in the F3V star (about 950\,km in the 500\,G
run). The depth of the depressions decreases strongly along our simulation
sequence towards cooler stars as well as with decreasing $B_0$. The dependence
on $B_0$ is mainly caused by the different size of the structures: in most
cases, the depressions are deeper in their central part than near their
boundaries (where the field strength is somewhat lower and where the side-wall
heating can result in a significant increase in temperature and thus
opacity). As the flux concentrations become larger at higher values of $B_0$,
their area becomes, on average, larger in relation to their
perimeter. Therefore, these ``boundary'' effects become less important at
larger $B_0$, resulting in a greater average depth of the Wilson
depressions. In the right panel of Fig.~\ref{fig:WDall}, the depth of the
Wilson depressions is given in terms of pressure scale height. Although the
pressure scale height decreases by more than one order of magnitude (from
$\sim500\,\mathrm{km}$ in F3V to $\sim 35\,\mathrm{km}$ in M2V, see Paper~I),
the decreasing trend of the depth of the depressions along the model sequence
still remains. The depth of the depressions ranges from $0.17\,H_p$ (M2V,
$B_0=20\,\mathrm{G}$) to $1.5\,H_p$ (F3V, $B_0=500\,\mathrm{G}$). Here, a
threshold of $B(\tau_{\mathrm{R}}=1)=750\,\mathrm{G}$ was applied to identify magnetic
flux concentrations. Using different threshold values does not influence the
results qualitatively, but has a quantitative impact, in particular for the
20\,G and 100\,G runs: with a threshold of 500\,G rather than 750\,G, the
depth of the Wilson depressions decreases by roughly 20\%, 15\%, and 5\% for
the 20\,G, 100\,G and 500\,G runs, respectively. With a threshold of 1000\,G
rather than 750\,G, the depth of the Wilson depression increases by roughly
15\%, 10\%, and 5\% for the 20\,G, 100\,G and 500\,G runs, respectively. The
small impact on the 500\,G runs confirms that the dependence on $B_0$ is
mostly due to the boundary effects discussed above.

%
%
\begin{table}
\caption{Magnetic pressure (in the flux concentrations) and external turbulent and thermal gas pressure at $z=0$ in the 500\,G runs in $\mathrm{dyn\,cm^{-2}}$.}\label{tab:pval}
\begin{tabular}{cccc}\hline\hline
Simulation & $p_{\mathrm{mag}}$ & $\rho \vel^2$ & $ p_{\mathrm{gas}}$ \\\hline
F3V & $7.41\times 10^4$ & $3.21\times 10^4$ & $7.65\times 10^4$\\
G2V & $9.39\times 10^4$ & $3.87\times 10^4$ & $1.46\times 10^5$\\
K0V & $1.10\times 10^5$ & $3.31\times 10^4$ & $2.28\times 10^5$\\
K5V & $1.23\times 10^5$ & $3.49\times 10^4$ & $3.15\times 10^5$\\
M0V & $1.47\times 10^5$ & $2.76\times 10^4$ & $5.11\times 10^5$\\
M2V & $1.51\times 10^5$ & $2.00\times 10^4$ & $5.99\times 10^5$\\\hline
\end{tabular}
\end{table}
%
%
\section{Balance of magnetic, thermal and turbulent pressures}\label{sec:pr_ba}
Owing to the Wilson depressions, the optical surface within the magnetic flux concentrations is below its average level $z=0$. As the magnetic field lines fan out in the upper parts of the
flux concentrations, the field strength is higher at the optical surface
(where it defines $B_{\mathrm{strong}}$) than at $z=0$
(cf.~Fig.~\ref{fig:WD}). The magnetic field strength $\langle
B(z=0)\rangle_{\mathrm{int}}$ in the magnetic flux concentrations
varies somewhat more strongly along our sequence of spectral types than $B_{\mathrm{strong}}$, but still
much less than one would expect if the convective collapse for thin flux tubes
\citep[cf.][]{Spruit76} were equally efficient for different stellar
parameters: this would yield $\langle B^2(z=0)
\rangle_{\mathrm{int}}/(8\pi)=p_{\mathrm{mag},\mathrm{int}}(z=0)\propto
p_{\mathrm{gas},\mathrm{ext}}(z=0)$, i.\,e. the magnetic pressure inside the
flux tubes (referred to as ``internal'' in what follows) should roughly scale
with the thermal pressure outside of the flux tubes (referred to as
``external'' in what follows), as the end products of an efficient convective collapse
are strongly evacuated magnetic structures which are roughly in pressure equilibrium with the
external gas pressure. Table~\ref{tab:pval} lists horizontal averages (at
$z=0$) of the internal magnetic pressure, the external gas
pressure, and the external ``turbulent pressure'' (twice the kinetic energy
density) for the 500\,G runs. Magnetic flux concentrations were
defined as the area where $B(z=0)>750\,\mathrm{G}$. The internal
magnetic pressure in the K- and particularly in the M-star simulations does
not reach equipartition with the external gas pressure, as it approximately
does for the Sun and the F3V star. In our model sequence (which roughly
follows the main sequence through the $T_{\mathrm{eff}}$-$\log g$ plane), the
decreasing efficiency of the convective collapse towards cooler stars is
roughly compensated by the increasing external gas pressure so that $B_{\mathrm{strong}}$ becomes almost
independent of the stellar type in our sequence.\par
%
\begin{figure*}
  \centering
  \includegraphics[width=7.2cm]{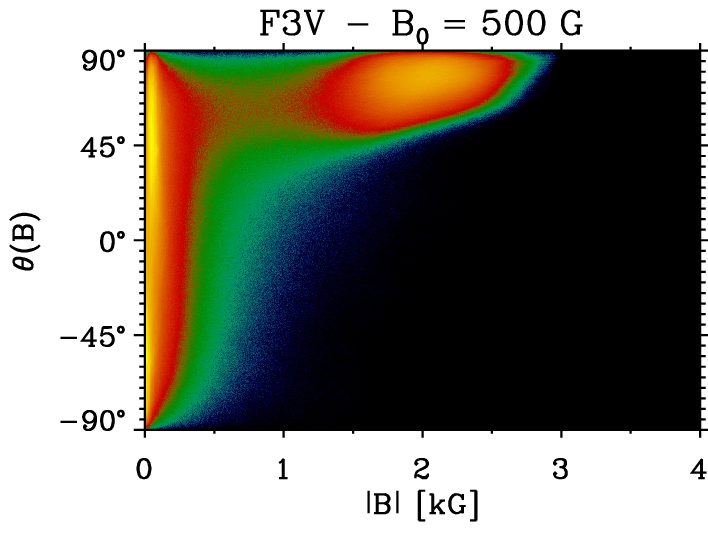}~~~\includegraphics[width=7.2cm]{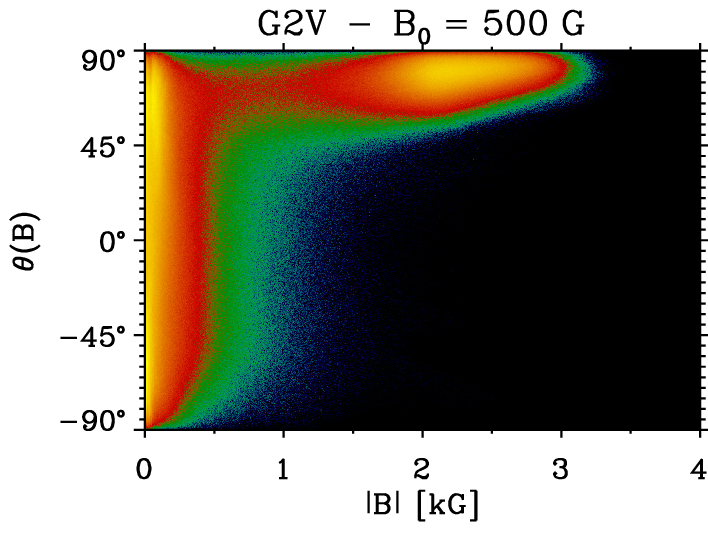}\\[1mm]
  \includegraphics[width=7.2cm]{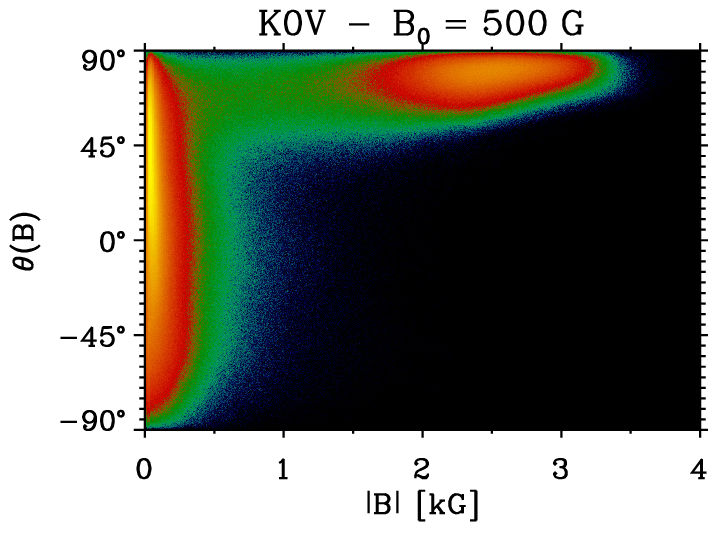}~~~\includegraphics[width=7.2cm]{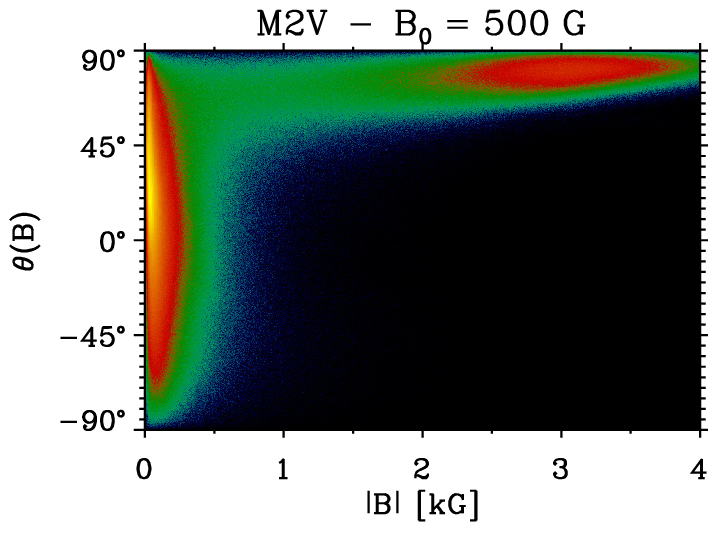}\\[1mm]
  \includegraphics[width=14.4cm]{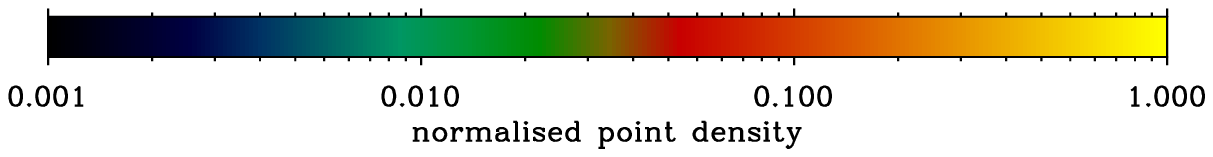}
\caption{2D histograms of the modulus of the field strength versus the
  inclination of the field (0$^{\circ}$ = horizontal; $\pm 90^{\circ}$ =
  vertical) for the 500\,G runs of the F3V, G2V, K0V and M2V stars.}\label{fig:Btheta}
\end{figure*}
Figure~\ref{fig:Btheta} shows 2D histograms of the magnetic field modulus
versus the inclination, $\theta$, of the field with respect to the surface ($\pm
90^{\circ}$ = vertical; $0^{\circ}$ = horizontal) in a layer covering half a
pressure scale height both below and above $z=0$. The magnetic field in the
near-surface layers falls into two components (cf. Fig.~\ref{fig:maghist}): in
regions with high field strength of a few kG (i.\,e. $\sim
B_{\mathrm{strong}}$) the field is mostly close to vertical while weak-field
regions show almost isotropic field inclinations with a weak tendency towards
positive inclinations (i.\,e. the field direction of the dominant
polarity). There is a dependence on spectral type: in the F3V-star
simulation, the inclinations in the strong-field regime are less concentrated
towards the vertical (with inclinations down to about $45^{\circ}$), while the field
inclination in the M2V star is much closer to vertical (always above
$60^{\circ}$) in this regime. In the latter simulation, the separation between
weak and strong field is also more pronounced. Along the model sequence there
is a smooth transition from the F-star-like to the M-star-like distribution as
illustrated by the other two spectral types shown in Fig.~\ref{fig:Btheta}. This is
related to the different balance of magnetic and turbulent
pressures (see Table~\ref{tab:pval}): in the F3V star, the ratio between
internal magnetic pressure and external turbulent pressure is about 2.5
whereas it is 30 in the M2V star. This renders the magnetic flux
concentrations in the F3V star much more susceptible to the surrounding plasma
motions, leading to a larger spread in field strength and orientation. In
contrast, the magnetic flux concentrations in the M-star simulations are quite
stationary and less strongly influenced by the surrounding convection.\par

\section{Heating of the upper photosphere}\label{sec:up_ph}
\begin{figure*}
 \centering
\includegraphics[width=7.1cm]{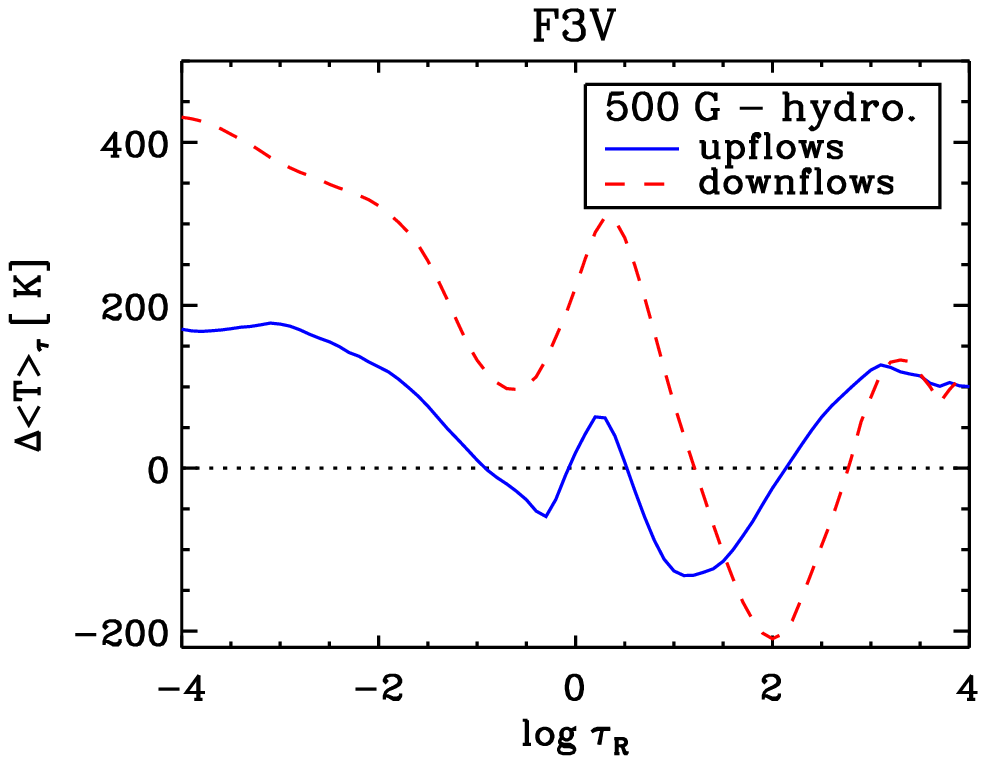}~\includegraphics[width=7.1cm]{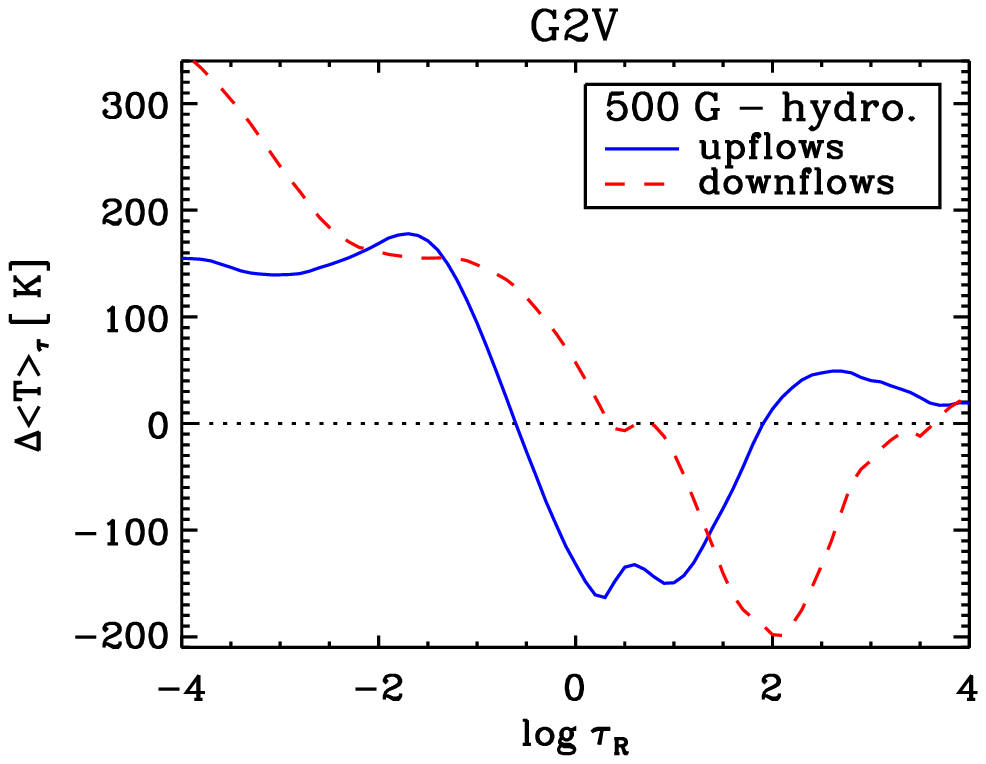}\\
\includegraphics[width=7.1cm]{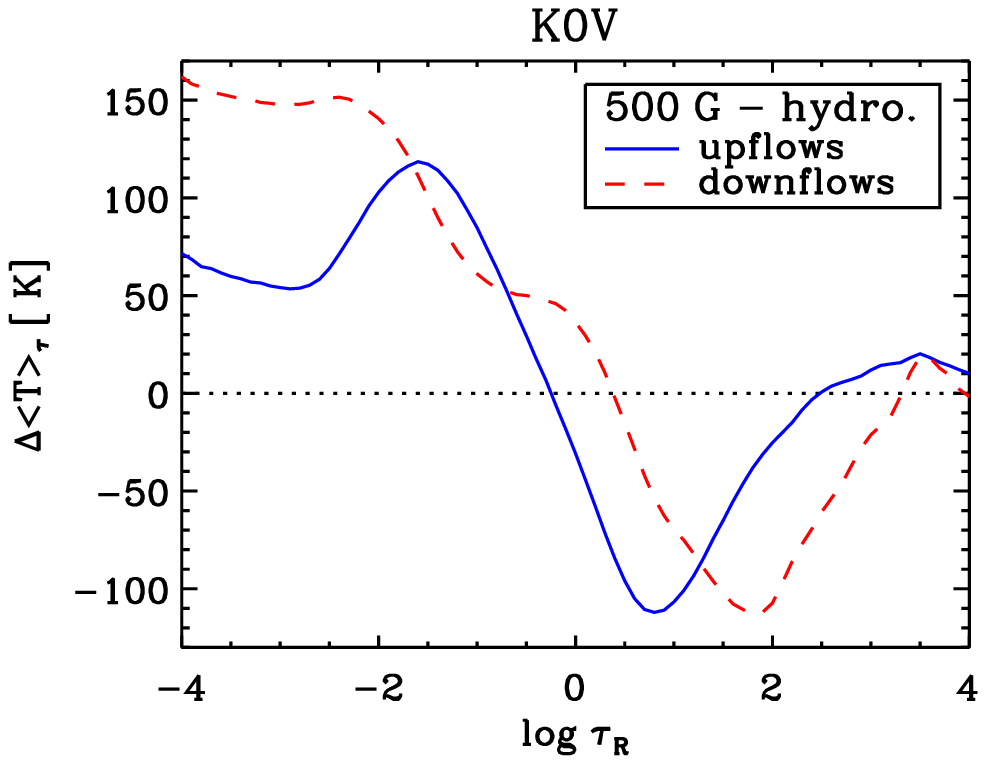}~\includegraphics[width=7.1cm]{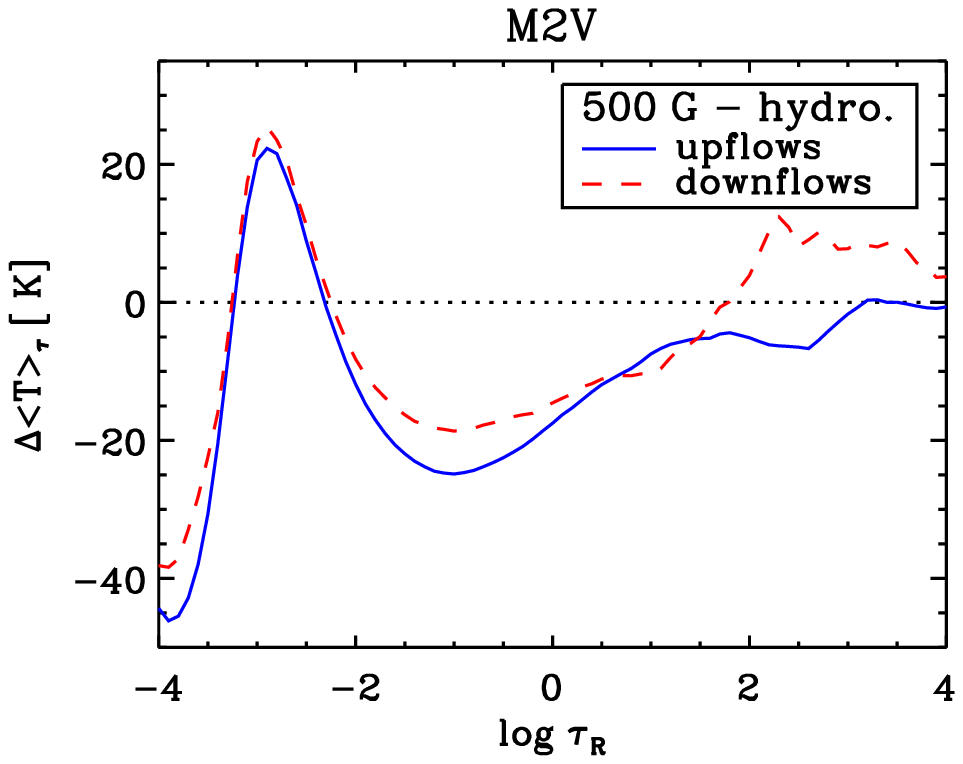}\\
\caption{Average temperature difference between the 500\,G and non-magnetic
  runs in up- and downflows for the F3V, G2V, K0V, and M2V
  simulations.}\label{fig:udf_T}
\end{figure*}
%
Figure~\ref{fig:udf_T} gives the difference in $\langle
T\rangle_{\tau}(\tau_{\mathrm{R}})$, the temperature averaged over
iso-$\tau_{\mathrm{R}}$ surfaces, between 500\,G runs and non-magnetic runs,
shown separately for up- and downflows. The local modification of the
temperature in the flux concentrations shows in the average temperature
profiles $\langle T\rangle_{\tau}(\tau_{\mathrm{R}})$. Interestingly, the
impact of the magnetic field on the temperature structure is qualitatively and
quantitatively similar in up- and downflows, although the magnetic flux
concentrations are mostly located in downflows: the temperature is somewhat
increased above the optical surface and decreased below the optical surface
with the exception of the M stars, where the impact is smaller in amplitude
and more complicated. The reduced temperature directly below the surface and
the enhanced temperature above the surface imply a reduced average temperature
gradient in the near-surface layers. This is a consequence of the somewhat
more efficient radiative energy transport across the optical surface. Owing to
the local depressions in the optical surface, the area of the radiating
surface is increased. As a consequence, the radiative flux and hence the
effective temperature of the atmospheres is somewhat higher in the magnetic
runs. As the M stars only have very shallow Wilson depressions, they do not
show a decreased temperature gradient around $\log\tau_{\mathrm{R}}=0$ in the
magnetic runs.\par In most cases, the middle and upper photosphere ($\log
\tau_{\mathrm{R}}\lesssim -2$) is warmer in the magnetic runs than in the
non-magnetic runs.
%
\begin{figure*}
\centering
\includegraphics[width=4.750cm]{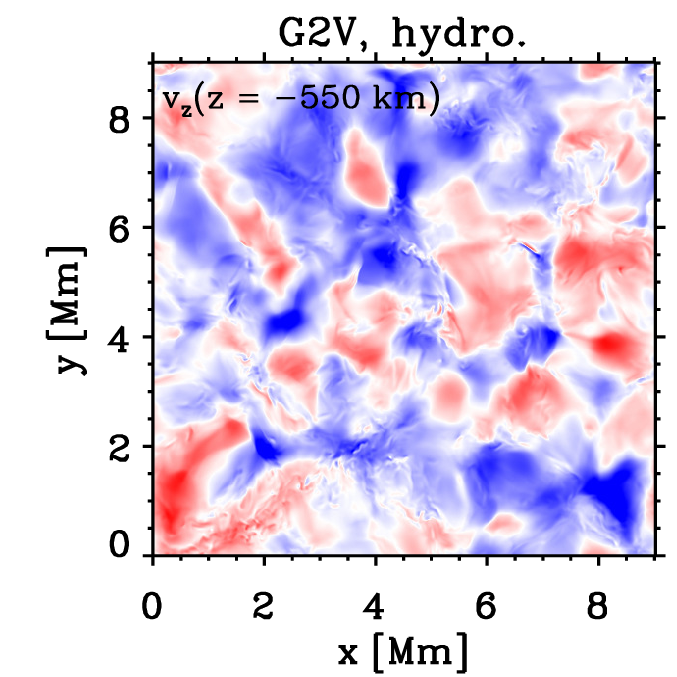}~\includegraphics[width=4.39cm]{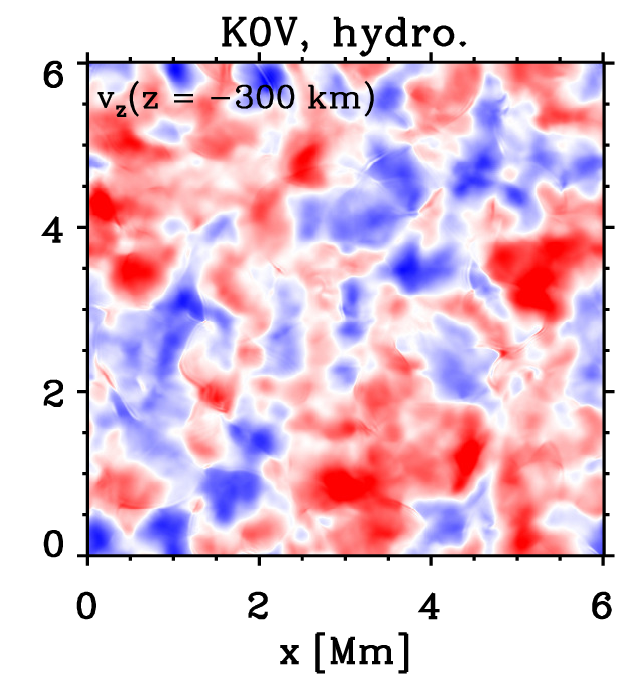}~\includegraphics[width=4.39cm]{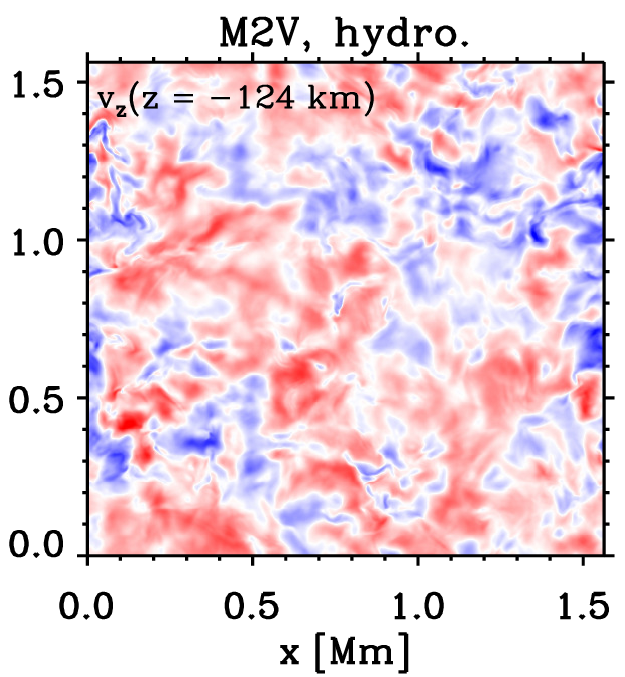}\\
\includegraphics[width=4.75cm]{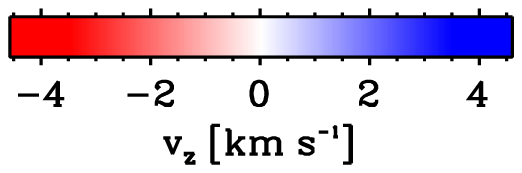}~\includegraphics[width=4.39cm]{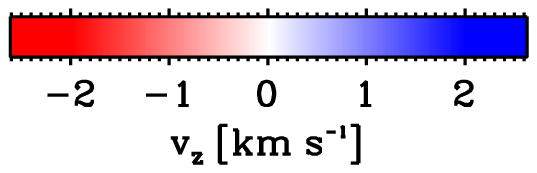}~\includegraphics[width=4.39cm]{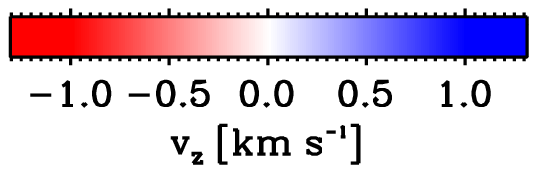}\\
\includegraphics[width=4.75cm]{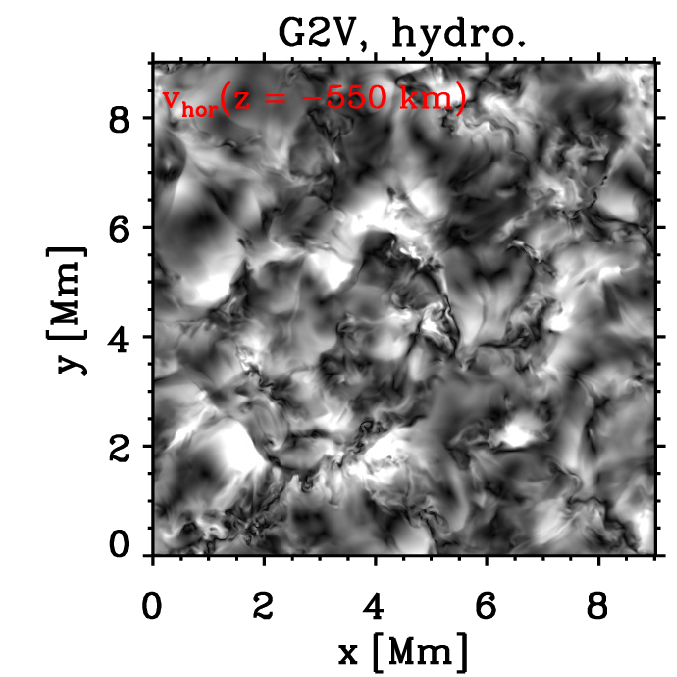}~\includegraphics[width=4.39cm]{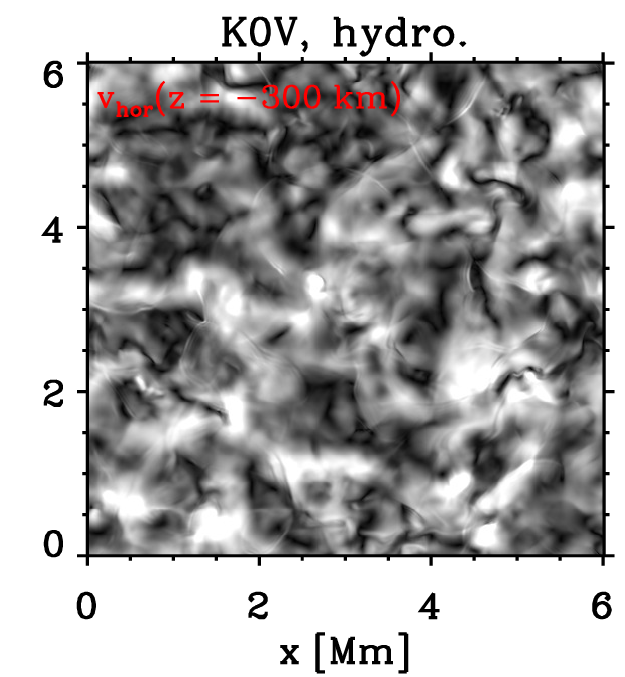}~\includegraphics[width=4.39cm]{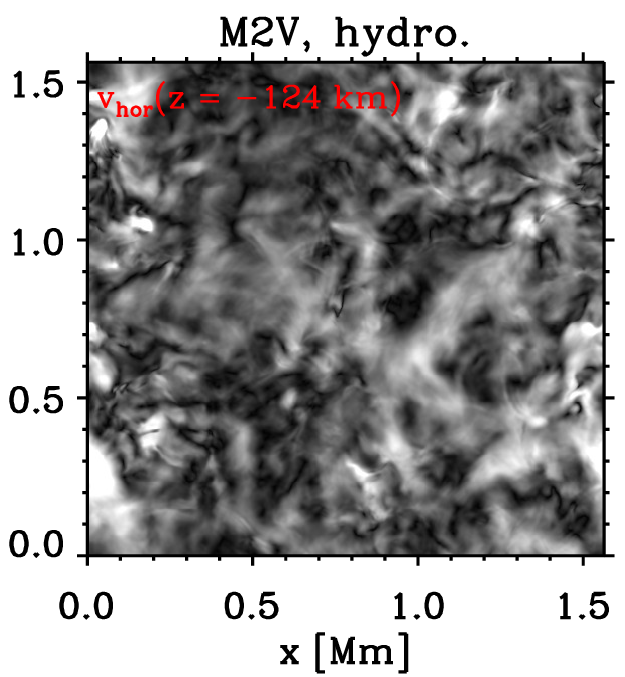}\\
\includegraphics[width=4.75cm]{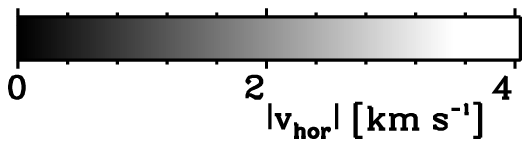}~\includegraphics[width=4.39cm]{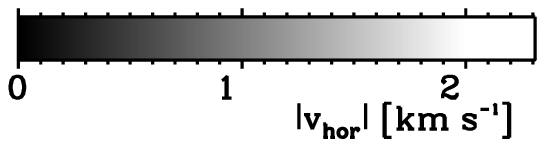}~\includegraphics[width=4.39cm]{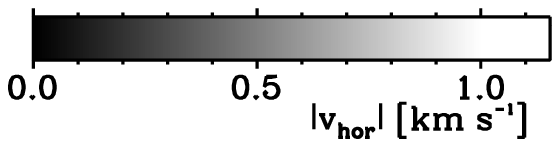}\\
\includegraphics[width=4.75cm]{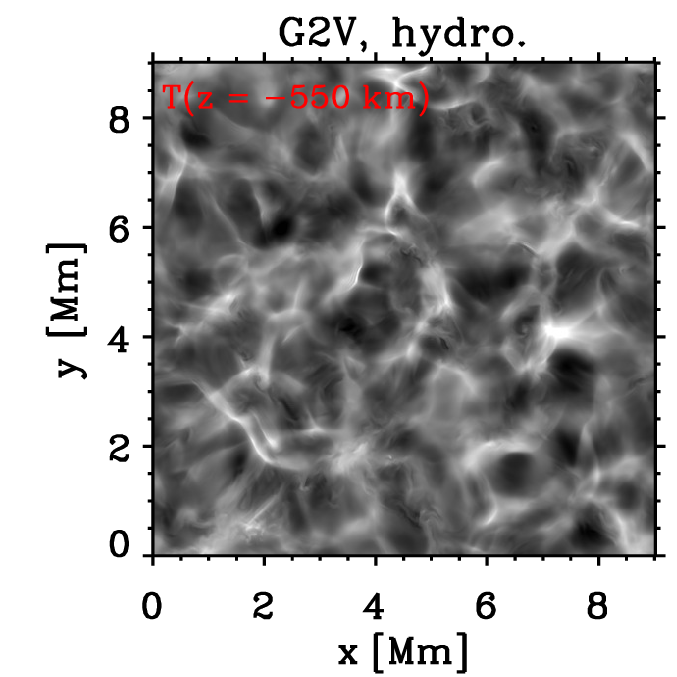}~\includegraphics[width=4.39cm]{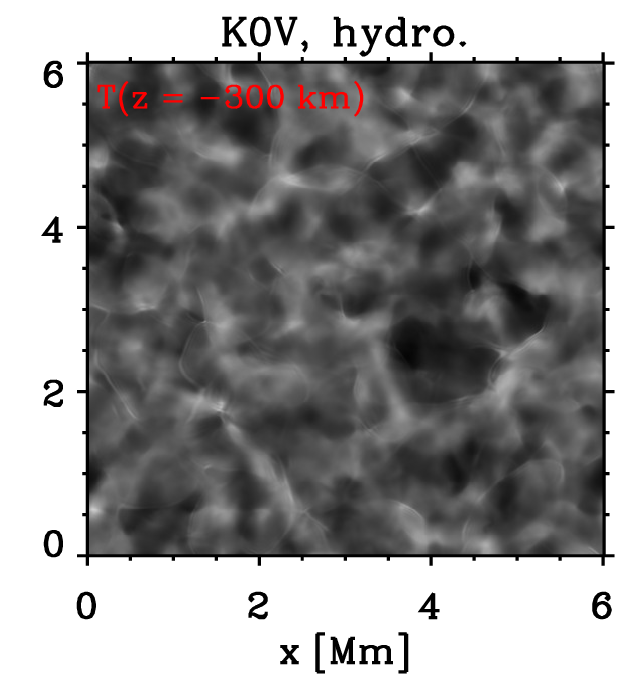}~\includegraphics[width=4.39cm]{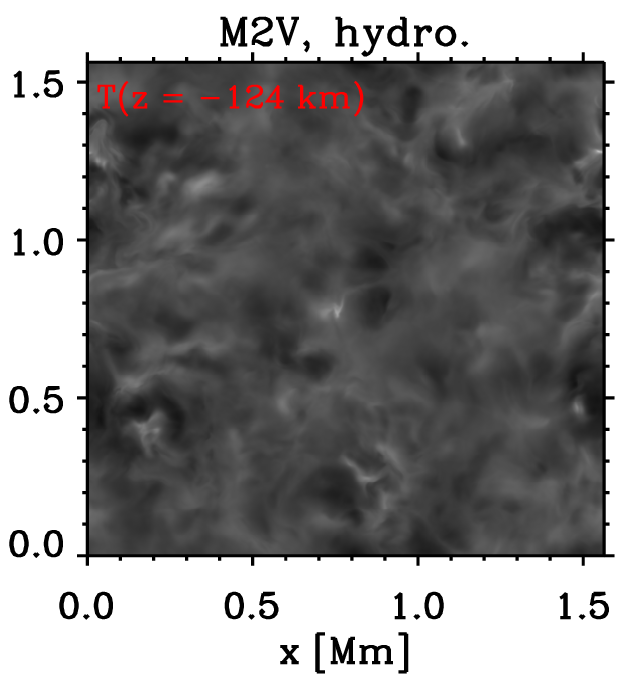}\\
\includegraphics[width=4.75cm]{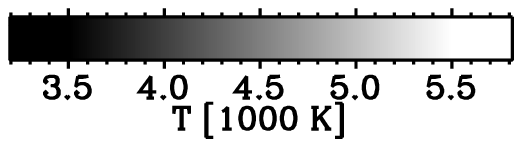}~\includegraphics[width=4.39cm]{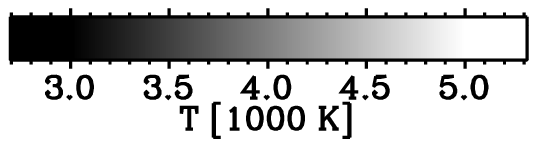}~\includegraphics[width=4.39cm]{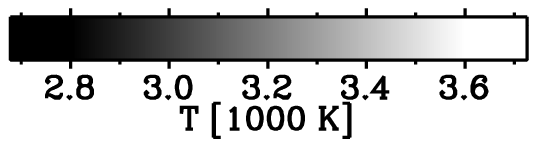}\\
\caption{Horizontal cuts through the boxes of the G2V, K0V, and M2V
  simulations ({\it from left to right}) without magnetic field. All cuts are
  situated such that $\langle p\rangle_z=0.01\,p_0$, i.\,e. about 4.6 pressure
  scale heights above the optical surface; the geometrical depth of this level
  is given in each panel. {\it From top to bottom:} Vertical velocity,
  $\vel_z$, horizontal speed,
  $|\vel_{\mathrm{hor}}|:=(\vel_x^2+\vel_y^2)^{1/2}$, and temperature, $T$ are
  given; see also Fig.~\ref{fig:heating_100G}.}\label{fig:heating_hydro}
\end{figure*}
%
\begin{figure*}
\centering
\includegraphics[width=4.750cm]{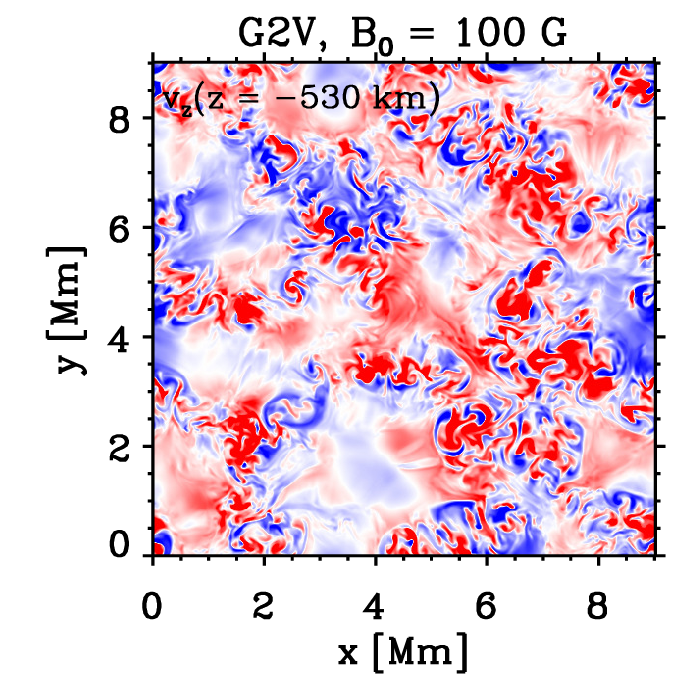}~\includegraphics[width=4.39cm]{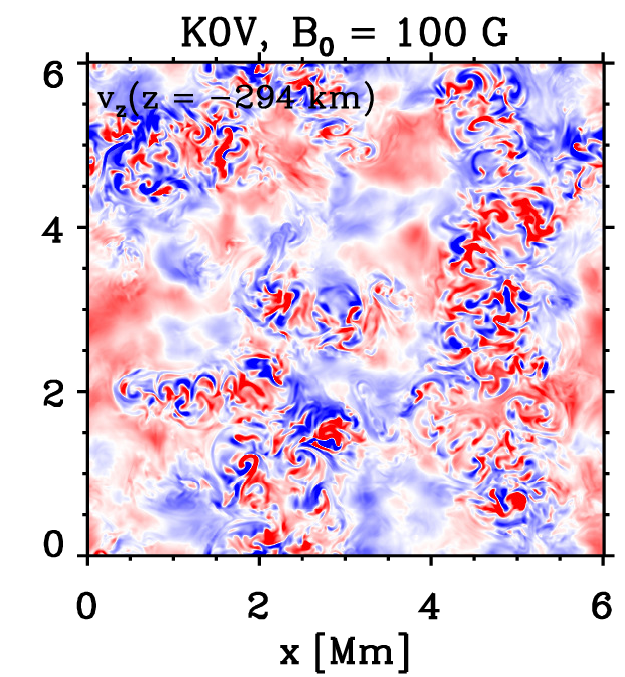}~\includegraphics[width=4.39cm]{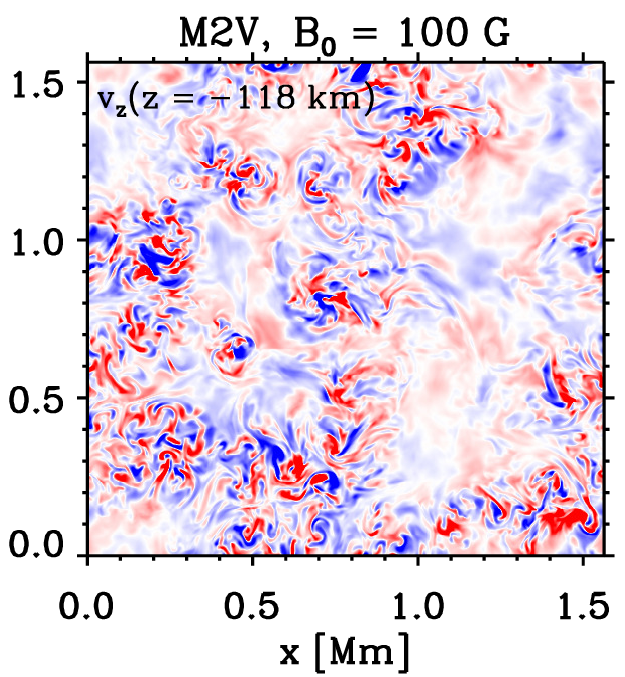}\\
\includegraphics[width=4.75cm]{hscale_vz_G2_1c.eps}~\includegraphics[width=4.39cm]{hscale_vz_K0_1c.eps}~\includegraphics[width=4.39cm]{hscale_vz_M2_1c.eps}\\
\includegraphics[width=4.75cm]{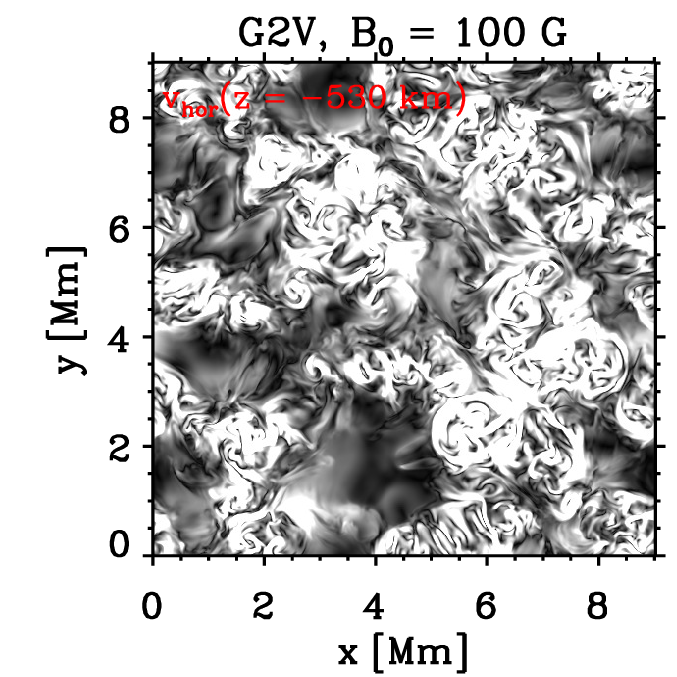}~\includegraphics[width=4.39cm]{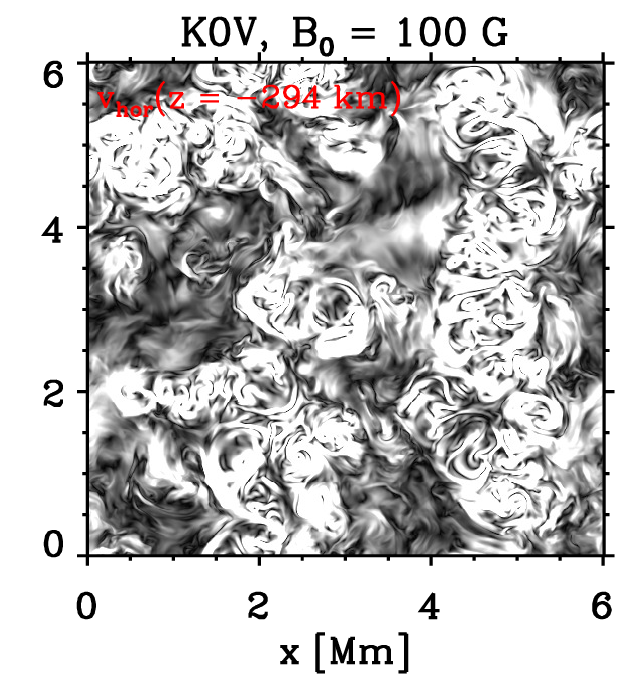}~\includegraphics[width=4.39cm]{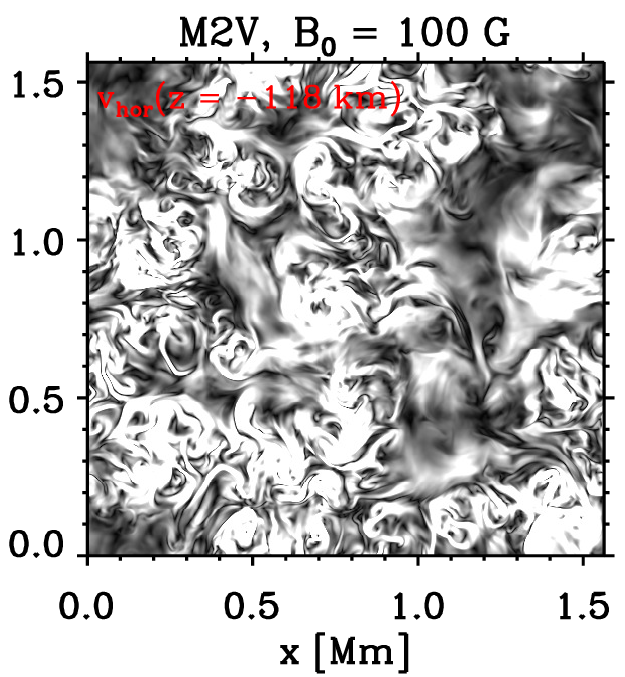}\\
\includegraphics[width=4.75cm]{hscale_vh_G2.eps}~\includegraphics[width=4.39cm]{hscale_vh_K0.eps}~\includegraphics[width=4.39cm]{hscale_vh_M2.eps}\\
\includegraphics[width=4.75cm]{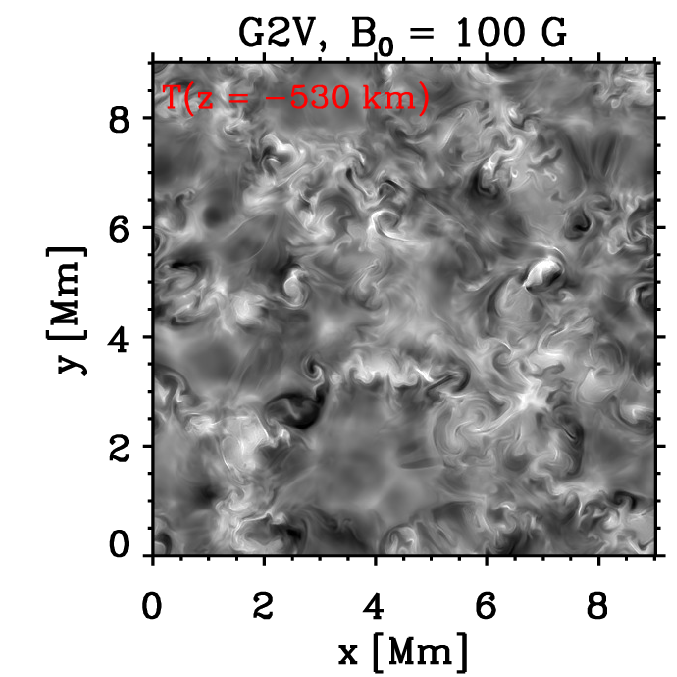}~\includegraphics[width=4.39cm]{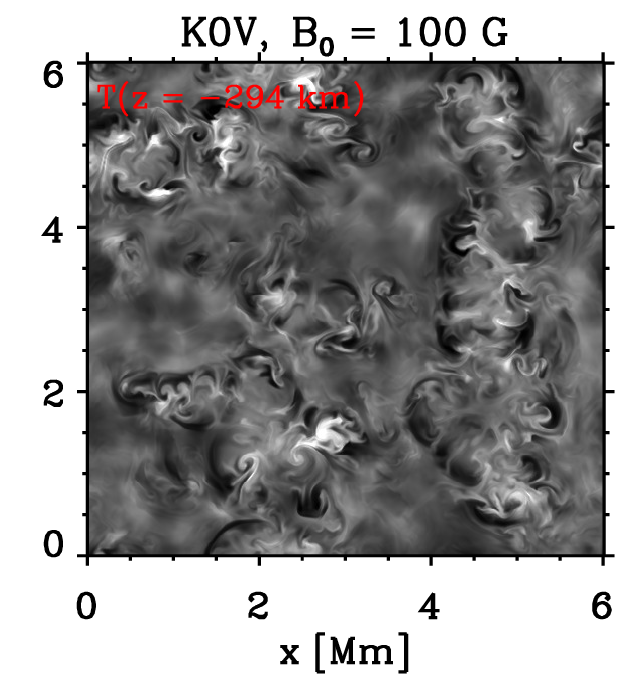}~\includegraphics[width=4.39cm]{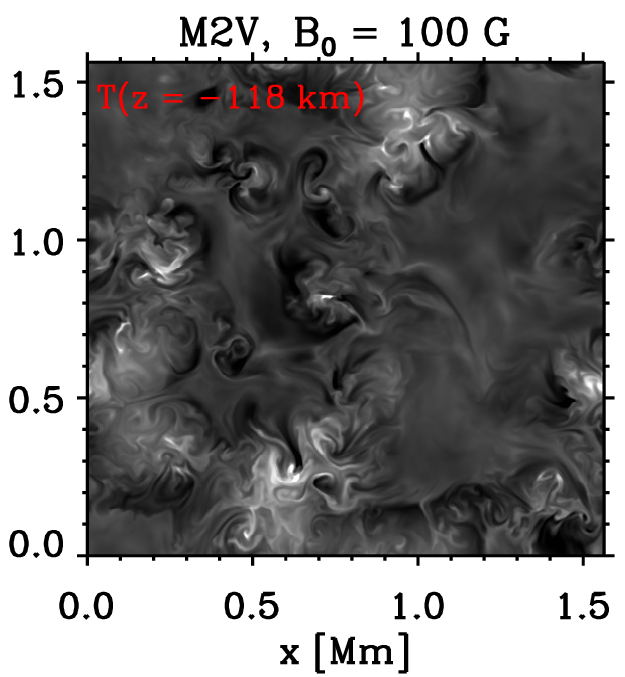}\\
\includegraphics[width=4.75cm]{hscale_T_G2.eps}~\includegraphics[width=4.39cm]{hscale_T_K0.eps}~\includegraphics[width=4.39cm]{hscale_T_M2.eps}\\
\hspace{13mm}\includegraphics[width=4.75cm]{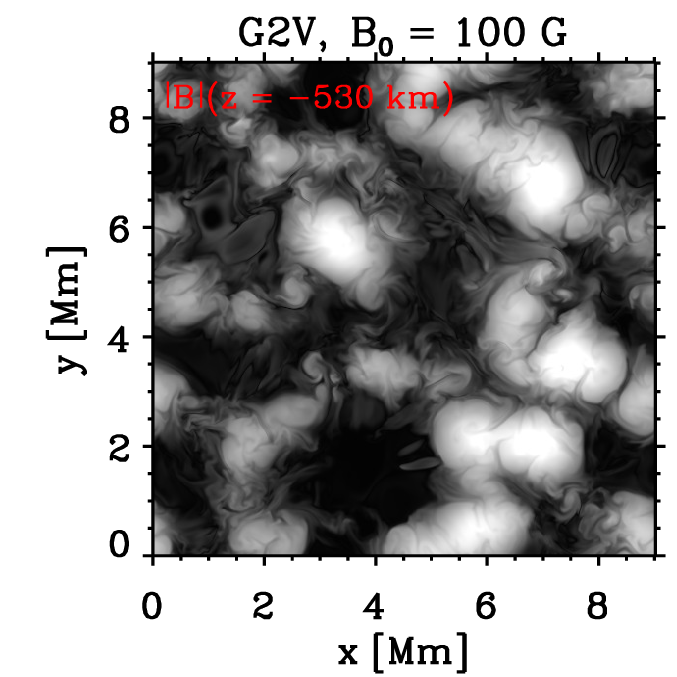}~\includegraphics[width=4.42cm]{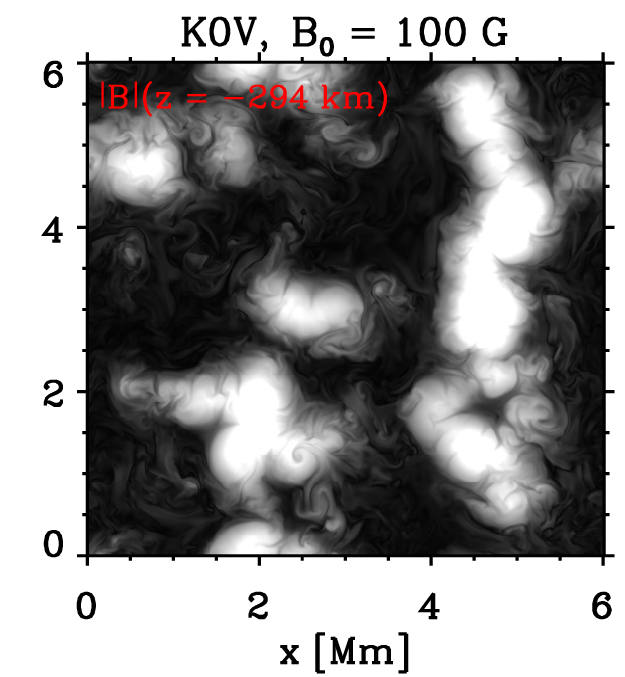}~\includegraphics[width=4.42cm]{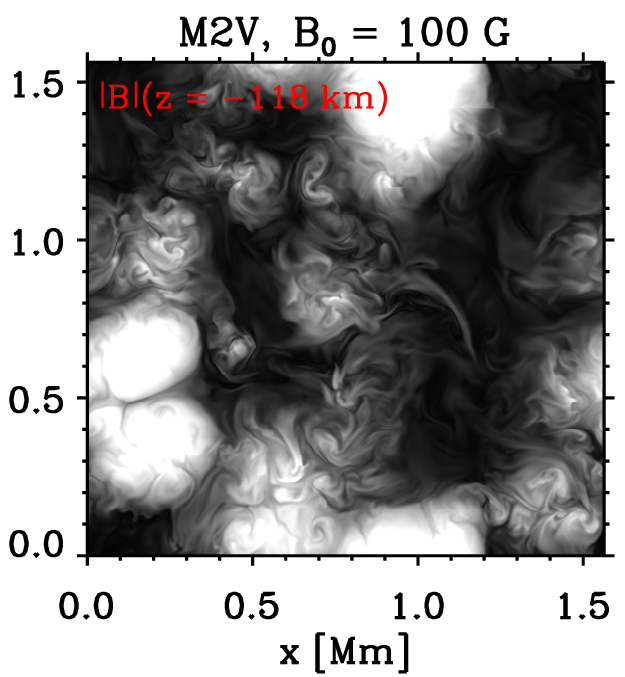}~\includegraphics[width=1.25cm]{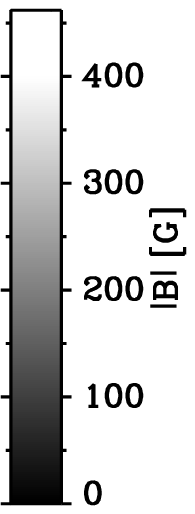}\\
\caption{The upper three rows give the same quantities as
  Fig.~\ref{fig:heating_hydro}, but for the 100\,G runs of the respective
  stellar model. The bottom row shows the horizontal field at the same
  height.}\label{fig:heating_100G}
\end{figure*}
%
While the heating of the lower photosphere above magnetic flux concentrations
is mainly due to radiation, the enhanced heating of the upper photosphere is
caused by enhanced viscous dissipation and Ohmic heating in regions of
substantial magnetic field. In the upper photosphere, the magnetic pressure
dominates the thermal pressure. The magnetic field couples the plasma at this
height to the magnetic flux concentrations below, which are subject to the
interaction with the convective flows. This excites magneto-acoustic waves as
well as torsional motions, which are then dissipated at greater heights. For
solar simulations this effect has been demonstrated by \citet{Rainer} and
\citet{Shelyag13}. Observational evidence for magneto-acoustic waves in
magnetic bright points in the Sun has been given by \citet{Shahin}.\par
Figures~\ref{fig:heating_hydro} and~\ref{fig:heating_100G} illustrate the
heating of the upper photosphere. The figures represent horizontal cuts through the
G2V-, K0V- and M2V-star simulations at a height of 4.6 pressure scale heights
above the mean level of the optical surface. Figure~\ref{fig:heating_hydro}
gives the non-magnetic runs of these stars, while Fig.~\ref{fig:heating_100G}
shows the 100\,G runs. The upper row of plots illustrates the vertical velocity at
that height. In the non-magnetic case (Fig.~\ref{fig:heating_hydro}), the
velocity amplitude is relatively low and the horizontal scale of the
structures is comparable to the granule size. In the 100\,G runs, small-scale
structures with high vertical velocities appear, which are associated with the magnetic structures (bottom row of Fig.~\ref{fig:heating_100G}). The second row of the figures shows
the modulus of the horizontal velocity at the same height.In the non-magnetic solar case, we find regions of high horizontal velocity that are usually
accompanied by a shock front at an interface between up- and downflows. In the
cooler stars, the horizontal velocities are relatively low and no shocks
appear. In the 100\,G run, the differences between spectral types almost
dissappear as all simulations show high-velocity horizontal flows with small
horizontal length scale at the same places where the small-scale vertical
velocity structures appear. The third row of the figures shows the temperature
at the same height. In the non-magnetic solar simulation, the shocks are
visible in the temperature as the localised viscous heating produces a high
temperature at the shock fronts. In the non-magnetic runs of the cooler stars,
the temperature fluctuates less at this level. In the magnetic simulations,
the small-scale high-velocity structures correspond to regions of high
temperature fluctuations (and in most cases regions of higher temperature). There
are some localised maxima where the temperature is more than 1000 K (in the
solar case even more than 2500 K) above the horizontal mean value. In
Fig.~\ref{fig:heating_100G}, the fourth row shows the magnetic field strength at
this height. The temperature maxima are all located in regions with high
magnetic field strength and often a significant gradient of the magnetic
field. Although the general structure of the magnetic field is much smoother on the
larger scales in this height, there is significant fine structure (e.\,g. current sheets and vortices),
which is responsible for the strong heating on small scales.

\section{Impact on convective flows}\label{sec:vel}
As described in Sect.~\ref{sec:magstr}, the velocity field is responsible for
the very inhomogeneous structure of the magnetic field in the first place:
horizontal outflows from the granules keep the magnetic flux in the downflow
regions, while the upflows become nearly field-free. Owing to this process
(in combination with the convective collapse), the field becomes locally very
strong (several kG) and, consequently, the magnetic pressure can locally
surpass the turbulent and thermal pressures (see Sect.~\ref{sec:pr_ba}). As a
consequence, the magnetic field reacts back on the velocity field.\par
%
\begin{figure*}
  \centering
  \includegraphics[width=4.75cm]{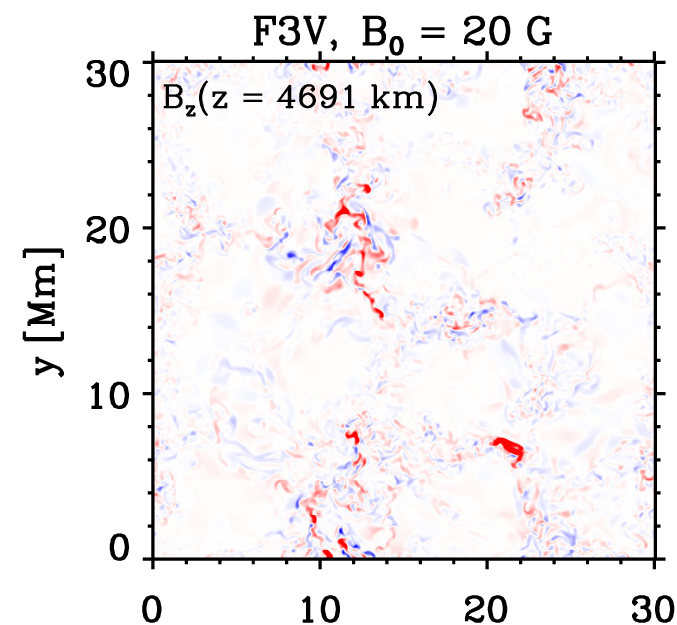}\includegraphics[width=4.42cm]{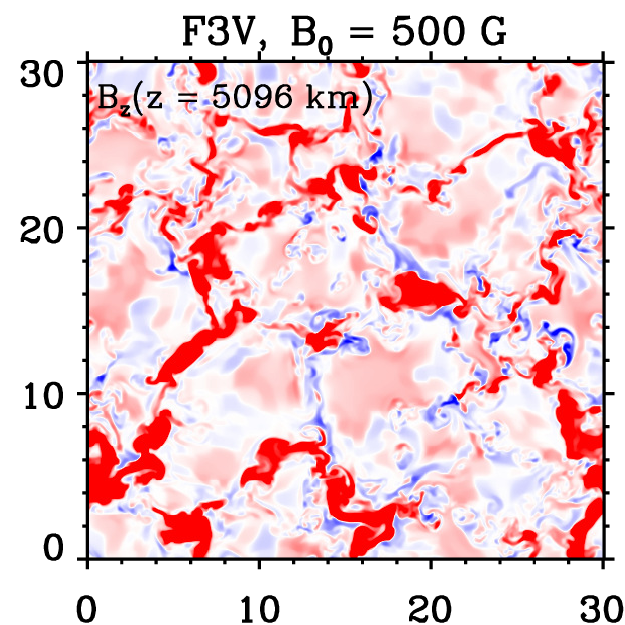}~~~~~~~~\includegraphics[width=4.42cm]{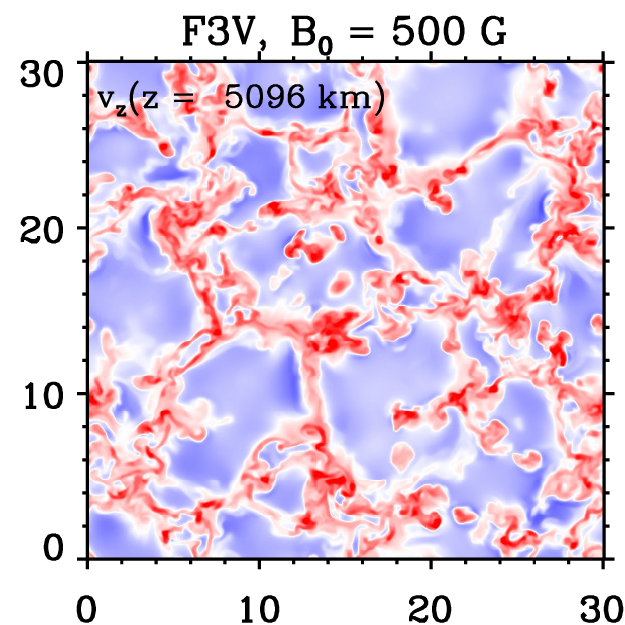}\includegraphics[width=1.45cm]{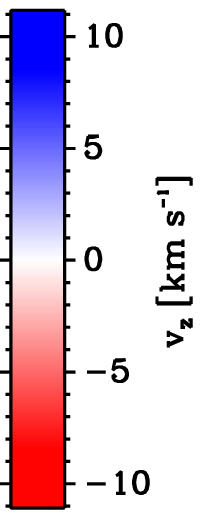}\\
  \includegraphics[width=4.75cm]{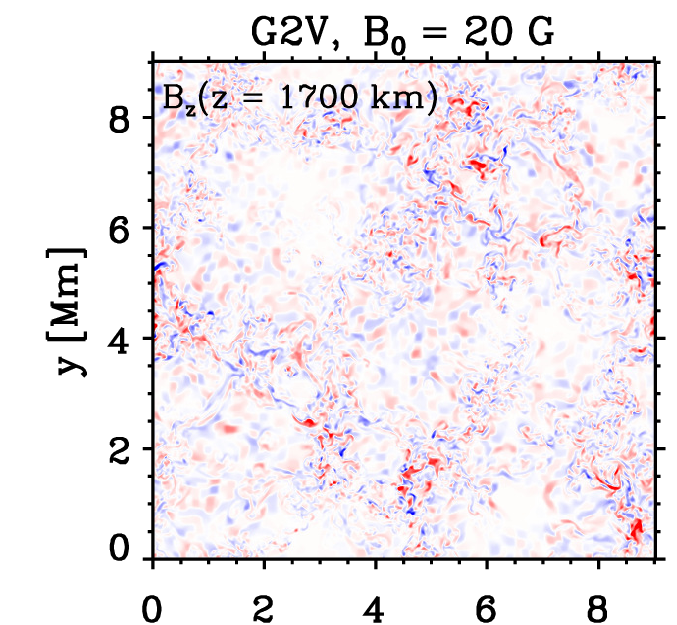}\includegraphics[width=4.42cm]{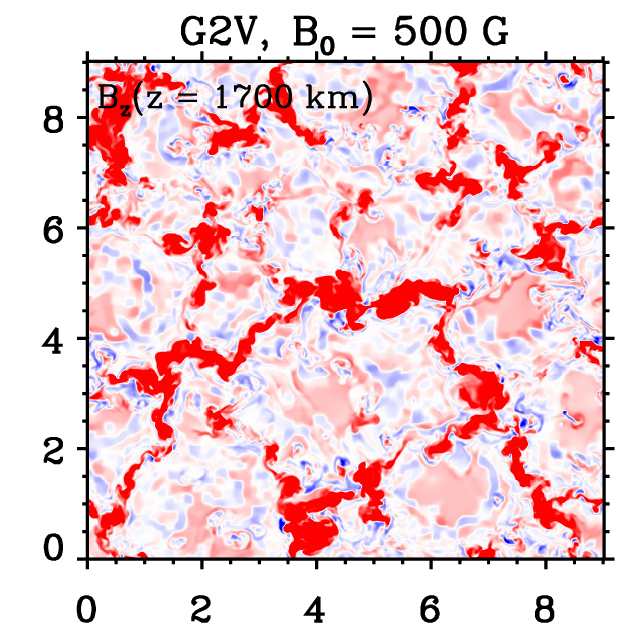}~~~~~~~~\includegraphics[width=4.42cm]{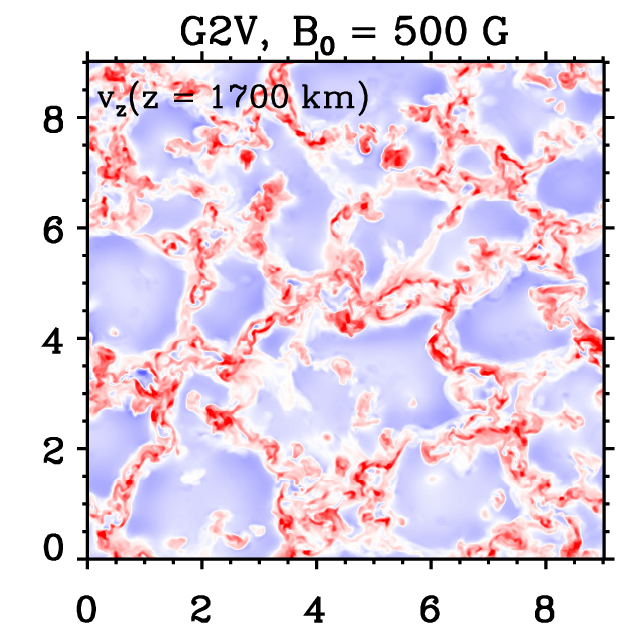}\includegraphics[width=1.45cm]{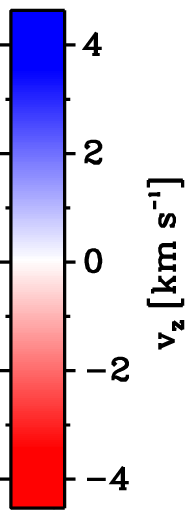}\\
  \includegraphics[width=4.75cm]{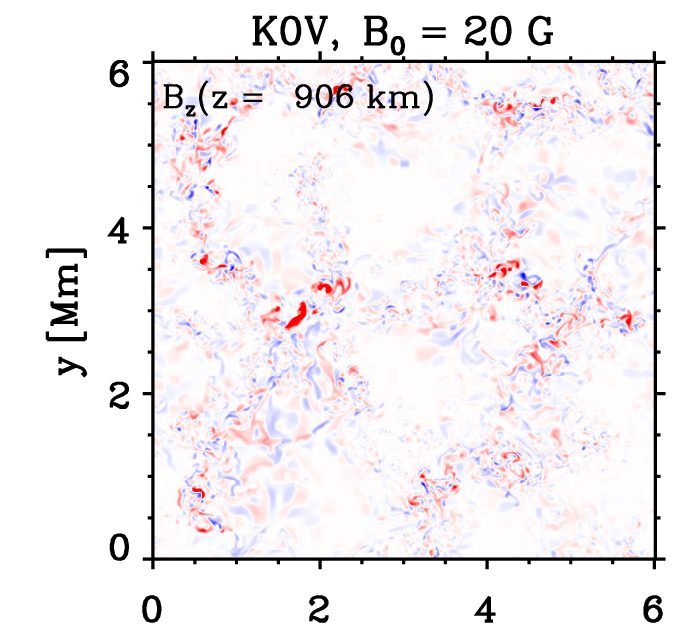}\includegraphics[width=4.42cm]{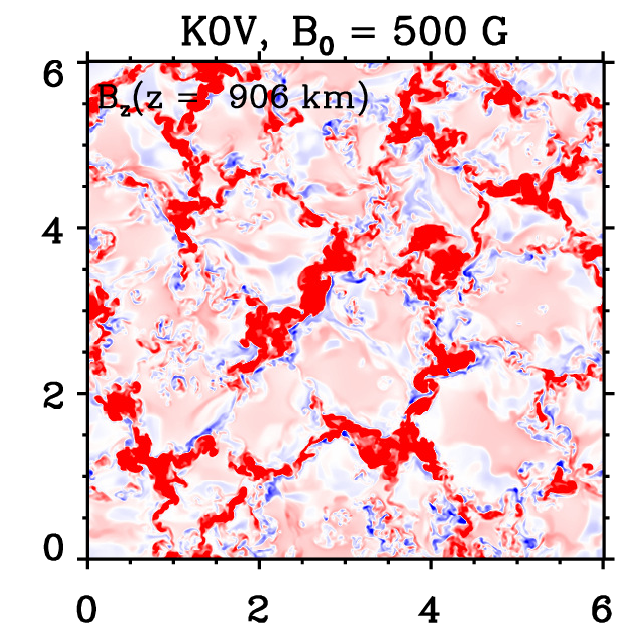}~~~~~~~~\includegraphics[width=4.42cm]{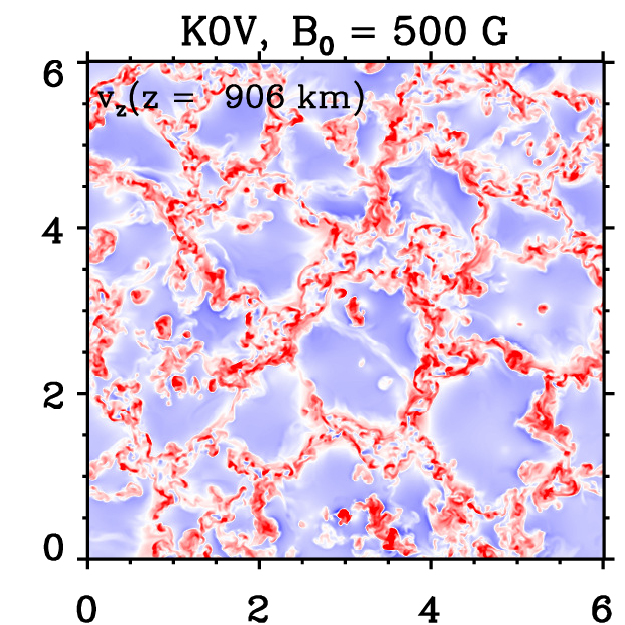}\includegraphics[width=1.45cm]{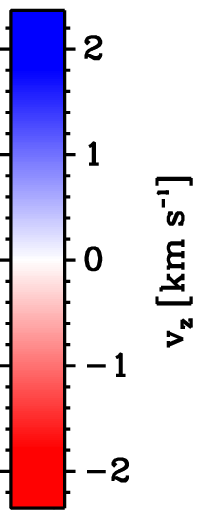}\\
  \includegraphics[width=4.75cm]{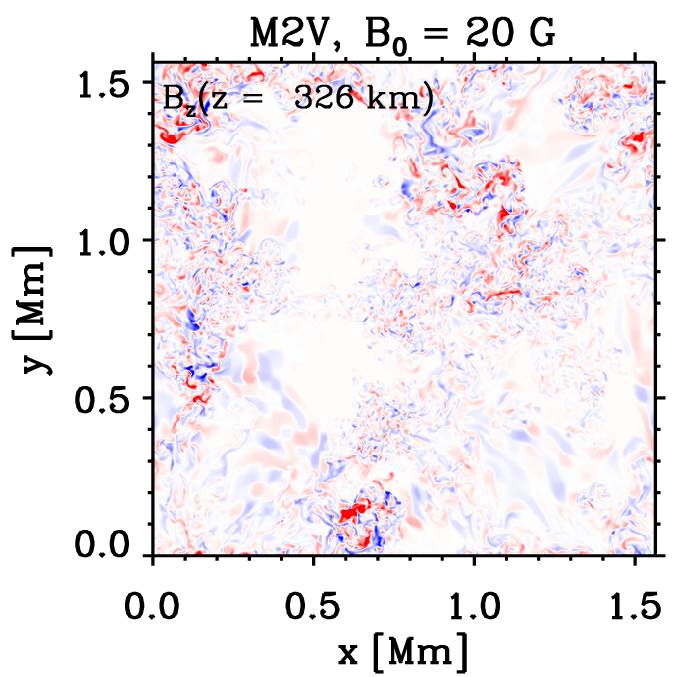}\includegraphics[width=4.42cm]{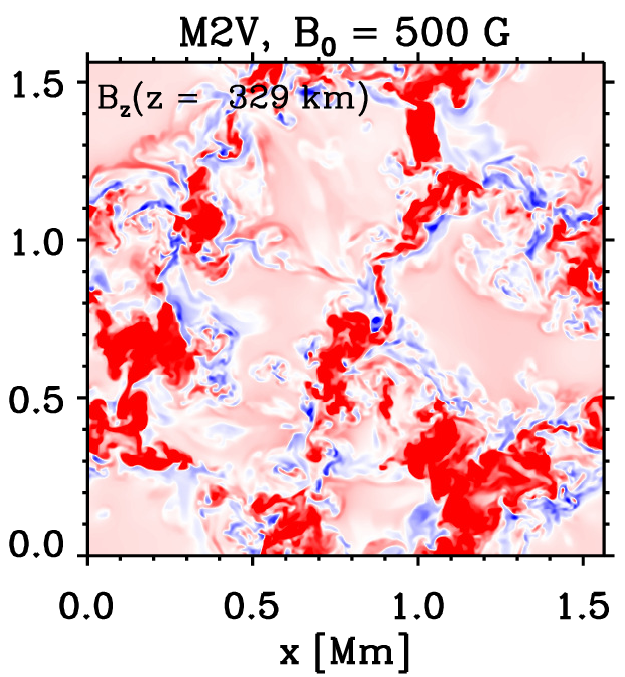}~~~~~~~~\includegraphics[width=4.42cm]{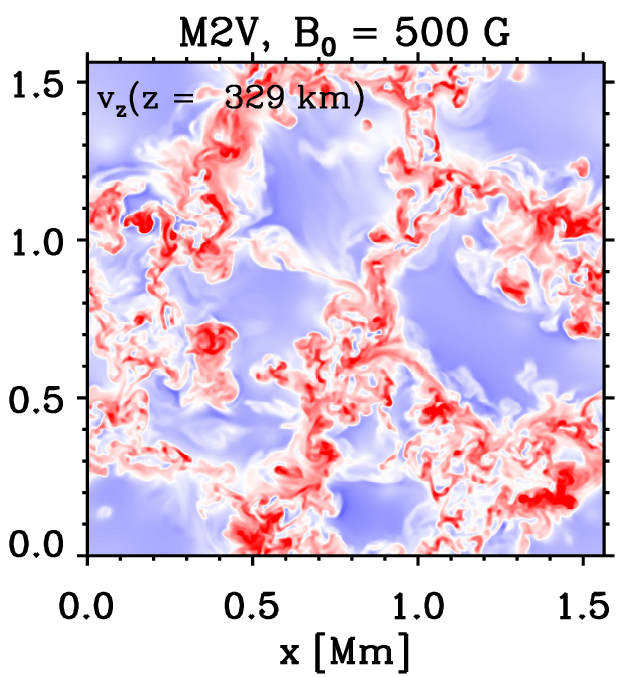}\includegraphics[width=1.45cm]{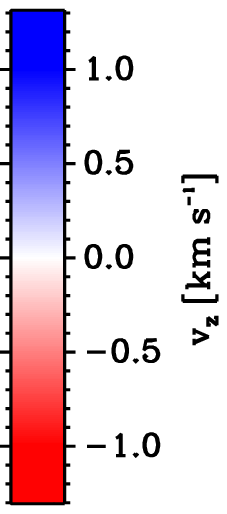}\\
  ~~~~\includegraphics[width=8.5cm]{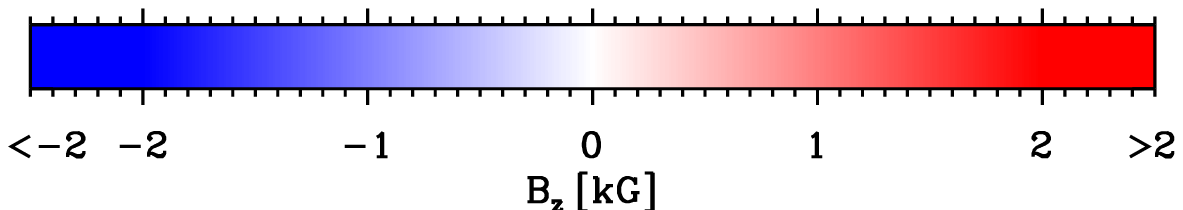}~~\hspace{6cm}

\caption{{\it Left two columns:} maps of the vertical magnetic field component of the 20\, and 500\,G runs of the F3V, G2V, K0V, and M2V simulations at a height
  corresponding to 4.6 pressure scale heights below the mean level of the
  optical surface. The geometrical depth $z$ of this level is specified in
  each panel. {\it Right column:} vertical component of the velocity of the 500\,G runs at the same depth.}\label{fig:magmap3}
\end{figure*}
%
Figure~\ref{fig:magmap3} displays maps of the magnetic field in a horizontal
cut located 4.6 pressure scale heights {\it below} the optical surface for the
20\,G and 500\,G runs of the F3V, G2V, K0V and M2V stars. There are patches of
both polarities. The dominating polarity (here shown in red) is concentrated
in the downflows. Surrounding these are some regions of the opposite polarity,
which are produced by field lines caught in the overturning motion at the
transition between up- and downflows. In the right column of the figure, the
vertical velocity at the same depth is shown for the 500\,G run. As the mass supply in the
magnetic flux concentrations from horizontal inflows is suppressed by the
strong field, the flux concentrations prevent the formation of strong
downdrafts below vertices of granules, which form in non-magnetic runs (see
Paper~I), and forces the downflows into a more network-like configuration
around the magnetic flux concentrations and below unmagnetised intergranular
lanes \citep[cf.][]{NS90}.\par
%
%
%
\begin{figure*}
\centering
\includegraphics[width=7.1cm]{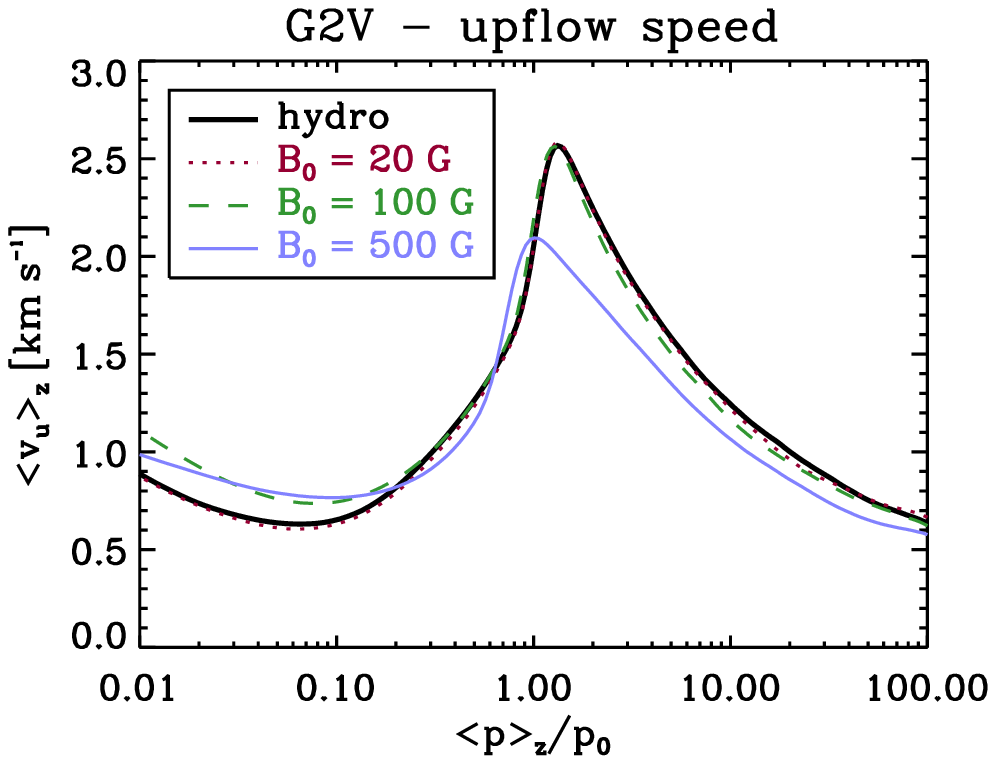}~\includegraphics[width=7.1cm]{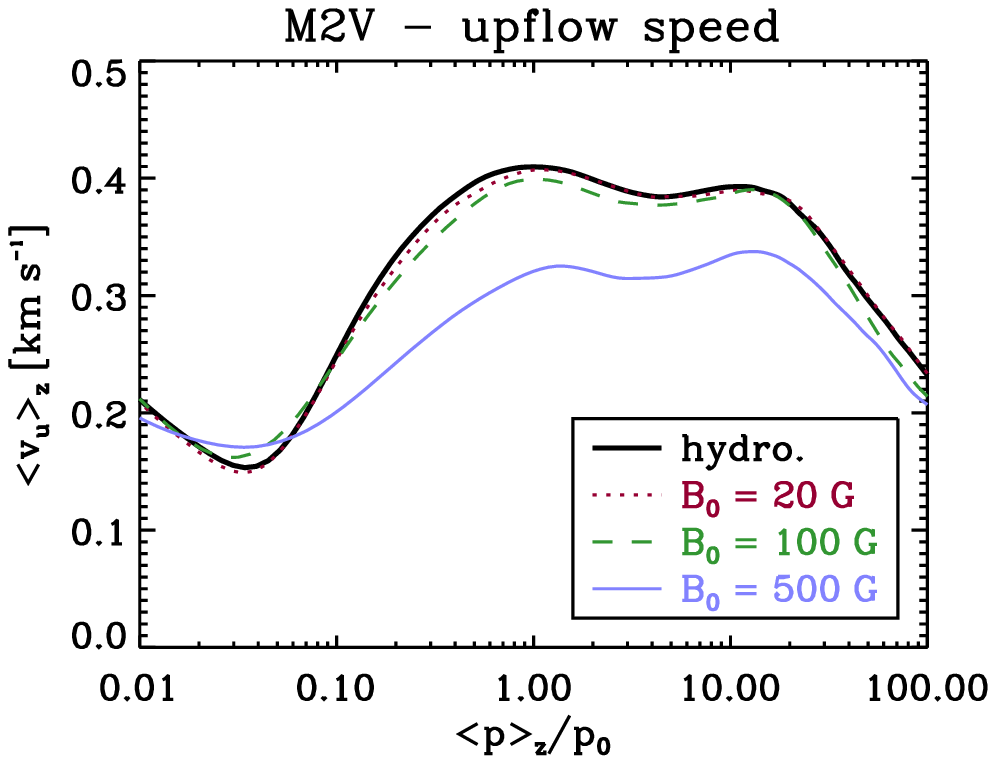}\\
\includegraphics[width=7.1cm]{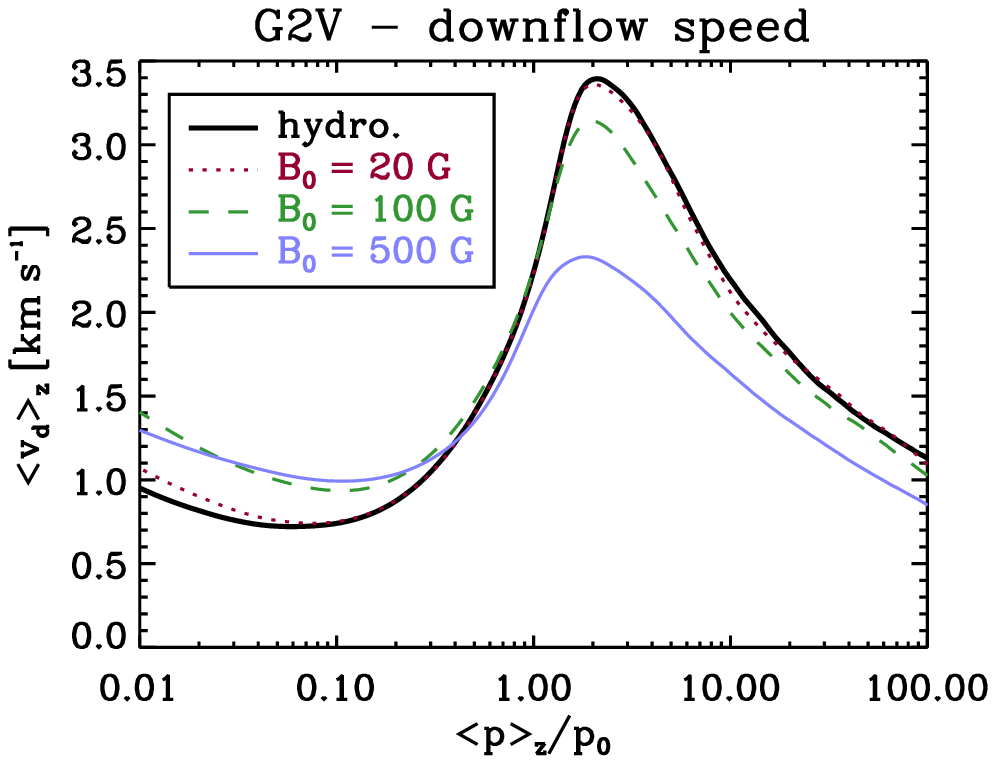}~\includegraphics[width=7.1cm]{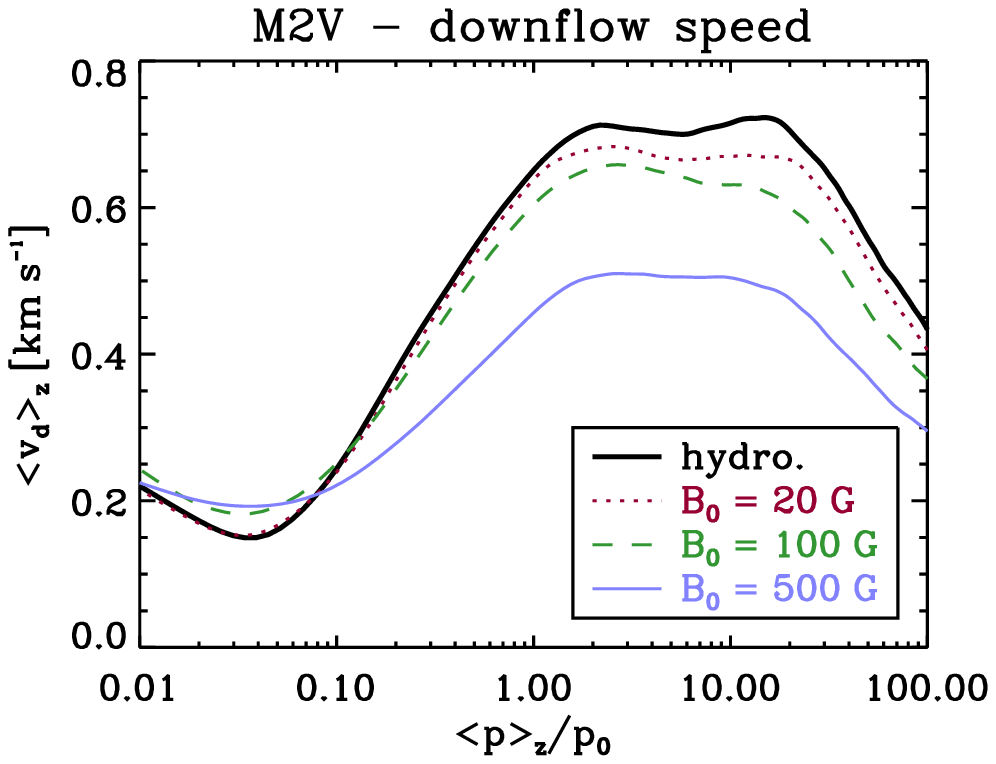}\\
\caption{Horizontal averages of the upflow ({\it upper panels}) and downflow ({\it lower panels}) speeds in the G2V ({\it left panels}) and M2V ({\it right panels}) simulations.}\label{fig:vud}
\end{figure*}
%
Figure~\ref{fig:vud} illustrates the effect of the magnetic field on the up-
and downflow speeds averaged over surfaces of constant geometrical depth
$z$. As the magnetic field is concentrated in downflows, the effect of the
magnetic field is stronger here. Below the surface, the downflows are slower
in the magnetic runs than in the non-magnetic runs. This effect becomes
stronger for increasing values of $B_0$. In the higher photosphere, the
magnetic runs often show a higher average downflow speed. This can be
interpreted as resulting from ``convective collapse'' events. The effect on the upflow speeds
is qualitatively similar but smaller. It turns out that an average only over
the weakly magnetised parts of the box (i.\,e. excluding the magnetic flux
concentrations), shows no significant impact on the upflow speed near the
optical surface whereas the downflow speed is still reduced by roughly 20\% in
the 500\,G runs. This shows that the effect on the upflows at the surface is
only due to the reduction of the area which is available for normal
convection, while the downflows are slower in the magnetic runs even in
non-magnetic parts of the box. This is possibly a consequence of the more
network-like structure of the downflows, which is more strongly affected by
dissipation of kinetic energy than the efficient downdrafts in the
non-magnetic runs.\par
We also find a change in the rms of the horizontal velocities, which are
smaller below the optical surface by 35 to 40\% in the 500\,G runs and up to
twice as high in the upper photosphere in the 100\,G runs compared to the
non-magnetic runs. The higher velocities in the upper photosphere
can be explained by the magnetic coupling to lower layers
(see Fig.~\ref{fig:heating_100G}).\par
%
\begin{figure*}
\centering
\includegraphics[width=7.2cm]{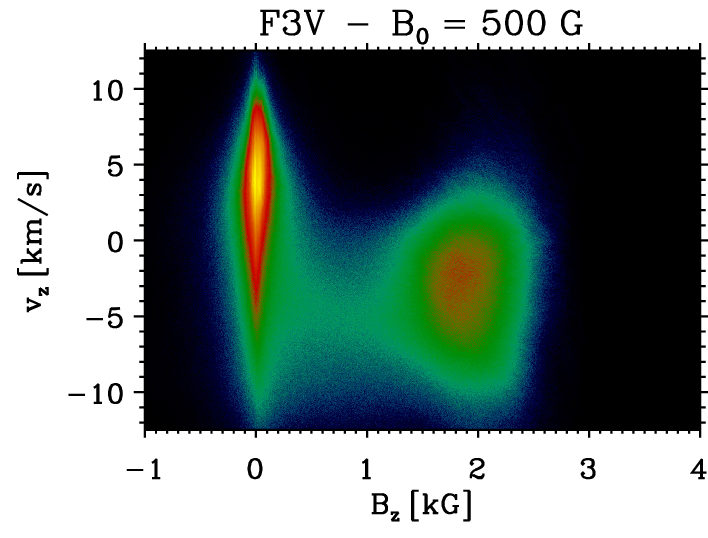}~\includegraphics[width=7.2cm]{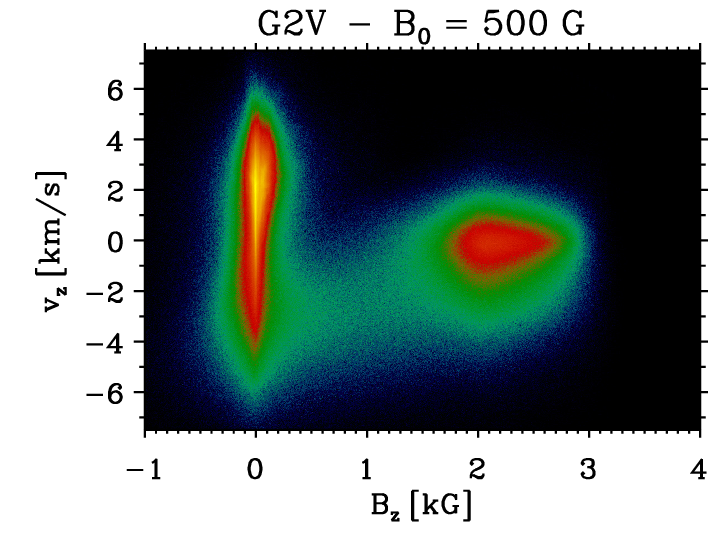}\\
\includegraphics[width=7.2cm]{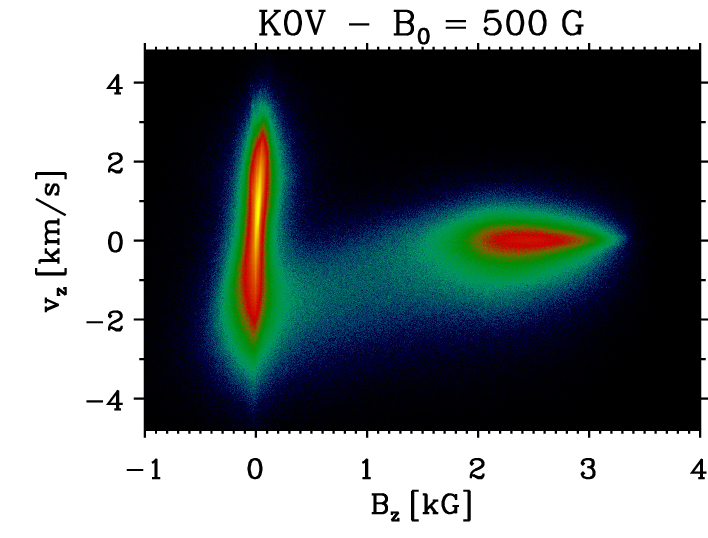}~\includegraphics[width=7.2cm]{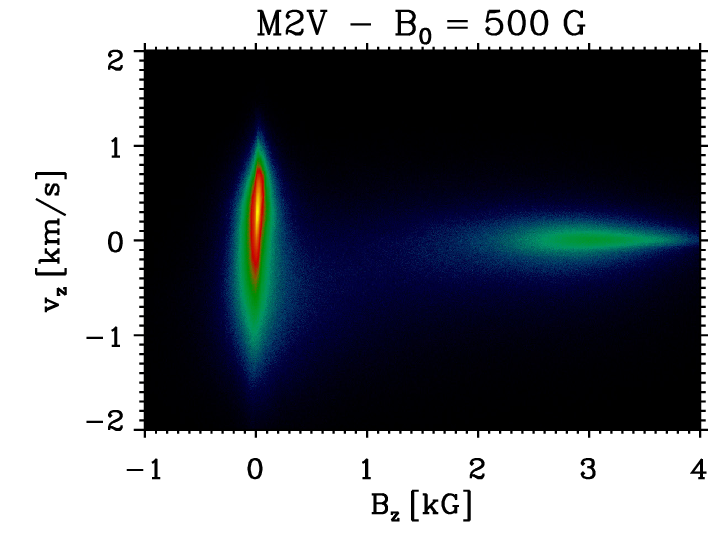}\\
\includegraphics[width=14.4cm]{scale_pdf.eps}
\caption{Joint histograms of the vertical components of the magnetic field and the flow velocity in a layer between $\langle z(p=p_0\,\exp(-0.5))\rangle$ and $\langle z(p=p_0\,\exp(+0.5))\rangle$, where $p_0=\langle p(\tau_{\mathrm{R}}=1)\rangle$ is the mean pressure at the optical surface.}\label{fig:pdf2}
\end{figure*}
%
Figure~\ref{fig:pdf2} shows 2D histograms of vertical velocity versus magnetic
field strength in a layer of a width of half a pressure scale height both
below and above $z=0$. The (kinematic) weak-field regime has a broad
distribution of vertical velocities as the upflows and a part of the downflows
are only weakly magnetised. At field strengths of up to a few 100\,G,
downflows are favoured. This is at the end of the kinematic regime, where the
magnetic field is passively advected into downflow regions and is not yet
strong enough to react back on the flows. These would include magnetic flux
concentrations during their (early) collapse phase. At high field strengths of
a few kG, the distribution is rather narrow around zero velocity with a
tendency towards downflows; however, the differences between spectral types
are most obvious here: while the Sun and especially the F3V star still have a
considerable spread in their velocities in the strong-field regime, magnetic
flux concentrations are essentially at rest in the K and M stars. This
probably is related to the different lifetimes of the magnetic flux
concentrations: while in the hotter stars, the individual magnetic flux
concentrations are short-lived and being constantly rearranged and fragmented
by the surrounding flows, the magnetic structures in the K- and M-dwarf
simulations are much longer-lived, so that most of them are in a stationary
phase at any given time. This can be understood in terms of the balance
between the turbulent pressure, the thermal gas pressure, and the magnetic
pressure in the flux concentrations (see Table~\ref{tab:pval}). Owing to the
much larger ratio of (internal) magnetic pressure to (external) turbulent
pressure in the magnetic flux concentrations of the cooler atmospheres of our
sequence, the disruptive effects of the flows on the magnetic flux
concentrations is much weaker.\par
%
As the only source of information on the magnetic field of stars are spectra
(more precisely, profiles of spectral lines sensitive to the Zeeman effect),
the relation between the magnetic and velocity fields becomes a problem:
active regions on a star have an rms velocity differing from quiet regions
resulting in a different velocity broadening of the spectral lines originating
from differently magnetised surface components. Moreover, in our F- and G-star
simulations, the magnetic field is mostly located in downflows, which adds a
non-zero Doppler shift to the magnetically broadened line component. However,
these effects are not taken into account at all when stellar magnetic fields
are spectroscopically determined. A detailed analysis of the effect of the
combined magnetic and velocity fields on a few spectral lines is given in
Paper~IV of this series.\par

\section{Conclusion}\label{sec:disc}

Solar observations show that a substantial fraction of the observed magnetic
flux is distributed in ``plage'' regions with a moderate average field
strength of up to a few 100\,G, which locally results in configurations very
close to the one discussed in this paper. In these regions, magnetic bright
points and faculae as well as (micro-)pores are observed, which are similar to
the structures visible in our and other comprehensive 3D RMHD simulations of
the Sun \citep[see e.\,g.][]{Carlsson04,Keller04,Cheung08}.\par
Here, we investigate how the appearance of moderately magnetised unipolar
regions changes for a sequence of cool main-sequence stars (including the
Sun). The major findings are:
\begin{itemize}
\item There is a lack of bright magnetic
  structures on M dwarfs \citep[cf.][]{CS16}. Owing to the more efficient convection (higher
  density near the optical surface, see Paper~I) in these stars, the
  superadiabatic peak is less pronounced. This reduces the efficiency of
  field amplification by a convective collapse of thin flux tubes
  \citep[cf.][]{Rajaguru02}. The resulting Wilson depressions are shallow
  so that heating by radiation from the side walls becomes rather inefficient.
\item The magnetic field strength in small-scale magnetic flux concentrations,
  $B_{\mathrm{strong}}$, is almost independent of spectral type and of the
  total magnetic flux available. Along the main-sequence the thermal pressure
  at the optical surface increases with decreasing effective temperature.
  This compensates the lower efficiency of the convective collapse.
\item For all spectral types considered, the upper photosphere is heated similarly by viscous and ohmic dissapation above
  small-scale magnetic field concentrations. The
  magnetic field dynamically couples the upper layers (where $p_{\mathrm{mag}}\gg
  p_{\mathrm{gas}}$) to the deeper layers (where $p_{\mathrm{mag}}\lesssim
  p_{\mathrm{gas}}$).
\item Convective flows are strongly modified within the magnetic flux
  concentrations. While the short-lived, smaller magnetic structures exhibit a
  strong downflow (convective collapse), the larger, long-lived structures are
  almost at rest as the inflows into the structures are blocked by the strong
  field.
\end{itemize}
We have shown that the spatial correlations between velocity field,
thermodynamic variables, and the magnetic field qualitatively and
quantitatively differ between stellar types. The impact of this on the
profiles of spectral lines in disc-integrated spectra, which are used for
magnetic field measurements, is investigated in Paper~IV of this series.
\begin{acknowledgements}
The authors acknowledge research funding by the Deutsche Forschungsgemeinschaft (DFG) under the grant SFB 963/1, project A16. AR has received research funding from the DFG under DFG 1664/9-2.
\end{acknowledgements}
\bibliography{paper3}
\end{document}